\documentclass[]{aa}

\usepackage{graphicx}
\usepackage{amsmath}

\usepackage{txfonts}

\usepackage{natbib}

\usepackage{hyperref}   
\usepackage{comment}

\hypersetup{
    bookmarks=true,         
    unicode=false,          
    colorlinks=true,       
    linkcolor=blue,          
    citecolor=blue,        
    filecolor=blue,      
    urlcolor=blue           
}

\begin{document}

\title{Impact of different approaches for computing rotating stellar models I. The solar metallicity case}

\author{Devesh Nandal\inst{1},
Georges Meynet\inst{1}, Sylvia Ekstr{\"o}m\inst{1}, Facundo D. Moyano\inst{1}, Patrick Eggenberger\inst{1}, Arthur Choplin\inst{2},
Cyril Georgy\inst{1}, Eoin Farrell\inst{1}, Andr\'e Maeder\inst{1}}

 \institute{$^1$D\'epartement d`Astronomie, Universit\'e de Gen\`eve, Chemin Pegasi 51, CH-1290 Versoix, Switzerland\\
 $^2$Institut d`Astronomie et d`Astrophysique, Université Libre de Bruxelles, Campus de la Plaine, Bruxelles, Belgium
}

 \authorrunning{Nandal et al.}
 \titlerunning{Impact of Rotation Prescriptions}


\abstract
{The physics of stellar rotation plays a crucial role in the evolution of stars, their final fate and the properties of compact remnants.}
{Diverse approaches have been adopted to incorporate the effects of rotation in stellar evolution models. This study seeks to explore the consequences of these various prescriptions for rotation on essential outputs of massive star models.}
{We compute a grid of 15 and 60 M$_{\odot}$ stellar evolution models with the Geneva Stellar Evolution Code (GENEC), accounting for both hydrodynamical and magnetic instabilities induced by rotation.}
{In both the 15 and 60 M$_{\odot}$ models, the choice of the vertical and horizontal diffusion coefficients for the non magnetic models strongly impacts the evolution of the chemical structure, but has a weak impact on the angular momentum transport and the rotational velocity of the core.
In the 15 M$_{\odot}$ models, the choice of diffusion coefficient impacts the convective core size during the core H-burning phase, whether the model begins core He-burning as a blue or red supergiant and the core mass at the end of He-burning.
In the 60 M$_{\odot}$ models, the evolution is dominated by mass loss and is less affected by the choice of diffusion coefficient. 
In the magnetic models, magnetic instability dominates the angular momentum transport and such models are found to be less mixed when compared to their rotating non-magnetic counterparts.
}
{Stellar models with the same initial mass, chemical composition, and rotation may exhibit diverse characteristics depending on the physics applied. By conducting thorough comparisons with observational features, we can ascertain which method(s) produce the most accurate results in different cases.}
\keywords{binaries:close-stars; stars: abundances; rotation;evolution
               }
\keywords{Stars: evolution -- Stars: rotation -- Stars: massive -- Stars: abundances }

\maketitle

\section{Introduction}
The impact of rotation on the evolution of stars has been the subject of extensive research \citep[e.g.][and references therein]{RMP2012}. The mechanical equilibrium of stars is altered by rotation, resulting in deformations \citep{Kipp1970,MM1997} and triggering instabilities that transport chemical elements and angular momentum \citep[see][for the equations of transport corresponding to various instabilities as they can be implemented in 1D stellar evolution models]{ES1981,Zahn1992, Maeder1998, Spruit2002}. Rotation influences mass loss due to line-driven stellar winds \citep{Maeder1999,Heger2000, Maeder2002, Vink2014, Bogo2021}, and can lead to mechanical mass loss when surface velocities reach the critical value where centrifugal acceleration balances gravity \citep[see, e.g.,][]{Krti2011,Georgy2013_winds}.

The inclusion of rotation in a star can result in a very different track in the Hertzsprung Russell (HR) diagram. On the modelling side, this phenomenon is well-illustrated by the case of chemically homogeneous evolution, which is triggered by efficient rotational mixing \citep{Maeder1987,Mandel2016,deMINK2016, Song2016}. The mass limits between different evolutionary scenarios are different depending on the initial rotation or initial angular momentum content at the zero-age main sequence (ZAMS). For instance, the lower mass limits for single stars evolving into the Wolf-Rayet phase \citep{MM2005} or undergoing a pair instability supernova depend on rotation \citep{Chatz2012, Marchant2020}. Nucleosynthesis in massive stars can be significantly influenced by their rotation, especially for certain isotopes. Notable examples include $^{26}$Al at solar metallicity \citep{Palacios2005, Brink2021,  Martinet2022, Brink2023}, $^{14}$N at very low metallicity \citep{Meynet2002}, and the s-process \citep{Pignatari2008, Frischknecht2016, Choplin2018, Limongi2018, Banerjee2019}. 
The pre-supernova stage is greatly affected by rotation where it influences the core collapse and the properties of the stellar remnant \citep{Hirschi2004, Hirschi2005,Limongi2018, Fields2022}. 
Rotation also appears as a key ingredient when exploring many topical astrophysical questions, such as the nature of the progenitors of the long Gamma Ray bursts \citep{HirschiGRB2005, YL2006}, the origin of the primary nitrogen in the early Universe \citep{Chiappini2006} and the spin and mass limits of stellar black holes and of neutron stars \citep{Heger2005, FullerMa2019, Griffiths2022, FullerW2022}.

In the last decades, many stellar evolution models both for single and binary stars accounting for the effects of rotation have been published \citep[see e.g.][]{Brott2011, Eks2012, Choi2016, Limongi2018, Renzo2021, Nguyen2022, Pauli2022, Renzo2023}. These models are based on different approaches for incorporating the effects of rotation and in particular, on the transport of the angular momentum and of the chemical species in the interior. As a result, models with a given initial mass, rotation and composition may produce significantly different outputs. The differences arise due the nature of the physical processes included and the way a given physical process is numerically implemented in the stellar evolution code. 

Rotating models for massive stars can be roughly classified into two main families: the non-magnetic and the magnetic rotating models. In this context, the term {\it magnetic} does not refer in the present case to the existence of a surface magnetic field \citep{GM2011, Zsolt2020}.
Instead, it refers to the action of a dynamo based on the magnetic Tayler instability \citep{Spruit2002}, hereafter called the Tayler-Spruit dynamo theory, or of another magnetic instability operating in the interior of a star. Examples of grids of non-magnetic rotating models are those of \citet{Eks2012, Choi2016, Limongi2018, Nguyen2022}. Examples of magnetic models can be found in \citet{Heger2005, Brott2011}. 

A first discussion of the impact of different prescriptions in the stellar code GENEC for rotation was presented in \citet{Meynet2013}. In this work we present a more extensive and detailed analysis. 
We study the outputs for non-magnetic and magnetic massive star models with masses 15 and 60 M$_{\odot}$ at solar metallicity obtained using different implementations for the various diffusion coefficients entering the theory (see Sect.~\ref{Sec:Coeff}). More precisely we study the implications of these different choices on the following model outputs: 
\begin{itemize}
\item The evolutionary tracks and lifetimes during the MS phase and the core helium burning phase.  
\item The changes of the surface composition and velocity.
\item The timescale for the crossing of the HR gap and the impact on the blue and red supergiant lifetimes.
\item The evolution of the internal angular velocity.
\item The properties the stellar cores at the end of the core He-burning phase.
\end{itemize}

We also discuss the impact of the different prescriptions on the 
boron depletion at the surface of massive rotating stars at the very beginning of the core hydrogen burning phase. The depletion of boron is a fascinating indication of shallow mixing that is probably associated with mixing processes within the star, rather than resulting from mass transfer in a close binary system \citep[see the discussion in][]{Fliegner1996}. As a result, this feature seems to really probe the physics of the mixing in a rather clean way \citep{Proffitt1999, Proffitt2001, Venn2002, Mendel2006, Frisch2010, Proffitt2016}. 
We discuss the effects of different prescriptions on the case of a 60 M$_\odot$ model at solar metallicity whose evolution is strongly impacted by mass loss through stellar winds. Finally, we also compare with magnetic models, a point that was not addressed in \citet{Meynet2013}. In an accompanying paper, we will study the impacts of these different implementations at very low metallicity (Nandal et al. in preparation).

We begin the paper with a brief discussion of the transport equations in rotating models in Sect. \ref{sec:equations} and the implementation of these equations in the stellar models in Sect.~\ref{Sec:Coeff}.  
In Sections~\ref{Sec:Results_nonmag} and \ref{Sec:Results_60}, we discuss the results for the 15 M$_\odot$ 60 M$_\odot$ and models accounting for hydrodynamical instabilities induced by rotation. 
The magnetic models for both the 15 and 60 M$_\odot$ models are presented in Sect.~\ref{Sec:Results_mag}.
In Sect.~\ref{Sec:Disc_Conc}, we present our conclusions.

\section{The transport equations} \label{sec:equations}

All the models considered in this work are assumed to have a near uniform rotation along an isobar (shellular rotating models). 
The homogenization of the horizontal angular velocity in a star is achieved by the presence a strong isobaric/horizontal turbulence, which also causes other physical quantities such as temperature and density to become uniform at a given level within the star \citep[see the discussion in][]{Zahn1992}. We assume this to be true because the restoring force (gravity) prevents strong turbulence in the vertical direction.

The transport of chemical species in the context of the shellular theory of rotation is governed by a purely diffusive equation \citep{Chaboyer1992} written as
\begin{eqnarray}
  \varrho{\partial X_i \over \partial t}\bigg|_{M_r} = {1 \over r^2} \, {\partial \over \partial r}
  \left (\varrho \, r^2 \, D_{\text{chem}} \, {\partial X_i \over \partial r}\right )  \; ,
\label{difx}
\end{eqnarray}
\noindent
where $X_i$ is the abundance in mass fraction of isotope $i$, and $D_{\text{chem}}$, 
the appropriate diffusion coefficient for the transport of the chemical elements\footnote{Equation (1) describes only the change of the abundance of an isotope at a given position due to the rotational diffusion. In the stellar models, two other processes, convection and nuclear reactions, are accounted for.}. The other symbols $\rho$, $r$, $M_r$ are the density, the radius and the lagrangian mass coordinate.

In a differentially rotating star, the evolution of the angular velocity $\Omega$ has to be followed at each level $r$, so that a full description of $\Omega(r,\, t)$ is available. 
In the case of shellular rotation, \citep{Zahn1992}, the equation describing the transport of angular momentum in Lagrangian form is
\begin{eqnarray} \label{eqn_ang_mom}
\varrho \, {\partial \over \partial t}(r^2 {\Omega})_{M_r}
={1 \over 5 \, r^2}{\partial \over \partial r}(\varrho \,  r^4 {\Omega} \,U_2(r))
+{1 \over r^2}{\partial \over \partial r}
\left(\varrho  \, D_{\text{ang}} \,  r^4 \,  {\partial {\Omega} \over \partial r} \right).
\label{eqn7}
\end{eqnarray}
\noindent
$U_2$ is the radial component of the velocity of the meridional circulation along the vertical direction, {\it i.e.} $U(r, \theta)=U_2(r) P_2(\cos\theta)$, where $P_2$ is the legendre polynomial of second order, and $D_{\text{ang}}$ is the appropriate diffusion coefficient for the transport of the angular momentum.

As explained in more detailed below, in this paper We discuss the outputs of numerical experiments, comparing
models that have the same set of assumptions and differing in the
algorithmic and numerical treatment of one particular aspect
(rotation).

\begin{table*}
\caption{Some properties of the models computed in the present work.}
\label{tab:1}       

\scriptsize{
\scalebox{0.9}{%
\begin{tabular}{lccccccccccccccccc}
\hline
Model &$D$      & $\overline \upsilon_{\rm eq}$ & $t_{\rm H}$ & $Y_{s-End H}$ & log${(N/H) \over (N/H)_{\rm ini}}$ & M$_{\rm End MS}$ & $t_{\rm He}$ &$t_{\rm blue}$ & $t_{\rm red}$ & $t_{\rm yel}$ & $t_{\rm red} \over t_{\rm blue}$ & $C^{12}_{c-End He}$  & $Y_{s-End He}$ & M$_{\rm EndHe}$ & M$_{\rm He}$ & M$_{\rm CO}$ & M$_{\rm rem}$ \\
   &   & km s$^{-1}$ & My & & & M$_\odot$& My & My& My & My &  & & M$_\odot$ & M$_\odot$ & M$_\odot$ &  M$_\odot$\\
\hline   
\multicolumn{18}{c}{} \\   
\multicolumn{18}{c}{15 M$_\odot$,$\upsilon_{\rm ini}/\upsilon_{\rm crit}=0.4$,$\upsilon_{\rm ini,surf} = 260 $ km s$^{-1}$} \\  
\multicolumn{18}{c}{ Z=0.014} \\  
\multicolumn{18}{c}{} \\
A & Ma97, Za92   & 188 & 13.72 & 0.290 &  0.62 &  14.70  & 1.37 & 0.04 & 1.23 & 0.10 & 27.89& 0.25 & 0.38& 11.20  & 5.08 & 3.01 & 1.66 \\
B &TZ97, Za92    & 196 & 13.70 & 0.268 &  0.24 &  14.74  & 1.13 & 0.00 & 1.13 & 0.00 & -    & 0.30 & 0.33& 13.37  & 5.83 & 3.46 & 1.77 \\
C & Ma97, Ma03   & 203 & 12.97 & 0.280 &  0.57 & 14.68   & 1.58 & 1.09 & 0.38 & 0.12 & 0.35 & 0.30 & 0.37&  10.31 & 4.56 & 2.57 & 1.54 \\
D & TZ97, Ma03   & 201 & 11.94 & 0.266 &  0.31 &  14.78  & 1.71 & 0.05 & 1.59 & 0.06 & 30.61& 0.18 & 0.33&  11.78 & 4.64 & 2.85 & 1.62 \\
E & Ma97, MZ04   & 172 & 12.69 & 0.276 &  0.53 &  14.75  & 1.55 & 1.05 & 0.38 & 0.11 & 0.36 & 0.30 & 0.36&  13.46 & 4.51 & 2.48 & 1.52 \\
F & TZ97, MZ04   & 179 & 11.40 & 0.267 &  0.44 &  14.80  & 1.72 & 1.27 & 0.33 & 0.11 & 0.26 & 0.25 & 0.33& 13.90  & 4.31 & 2.87 & 1.62 \\
C* & Ma97, Ma03  & 199 & 12.44 & 0.271 &  0.43 &  14.75  & 1.22 & 0.00 & 1.21 & 0.01 & 6040 & 0.34 & 0.31& 13.69  & 4.85 & 2.72 & 1.58 \\
G & Non-Rotating & 0.00& 11.26 & 0.231 &  0.42 &  14.80  & 1.37 & 0.37 & 1.11 & 0.00 & 2.98 & 0.30 & 0.30& 13.43  & 5.19 & 3.10 & 1.68 \\
\multicolumn{18}{c}{} \\
\multicolumn{18}{c}{60 M$_\odot$, $\upsilon_{\rm ini}/\upsilon_{\rm crit}=0.4$,$\upsilon_{\rm ini,surf} = 340$ km s$^{-1}$} \\  
\multicolumn{18}{c}{ Z=0.014} \\  
\multicolumn{18}{c}{} \\
A & Ma97, Za92  & 175   & 4.55 & 0.808  &  1.682  &  38.58  & 0.36  &   0.36 & 0.00 & 0.00 & 0.00 & 0.20 & 0.27  &  17.54   & 17.54 & 16.93 & 5.23 \\
B & TZ97, Za92 & 174    & 4.43 & 0.250  &  1.681  &  39.15  & 0.37  &   0.37 & 0.00 & 0.00 & 0.00 & 0.20 &0.28  &  15.88   & 15.87 & 15.23 & 4.71 \\
C & Ma97, Ma03 & 173    & 4.68 & 0.904  &  2.020  &  38.08  & 0.35  &   0.35 & 0.00 & 0.00 & 0.00 & 0.19 & 0.28  &  19.60   & 19.59 & 18.82 & 5.80  \\
D & TZ97, Ma03 & 170    & 4.60 & 0.252  &  2.121  &  38.30  & 0.35  &   0.35 & 0.00 & 0.00 & 0.00 & 0.19 & 0.28  &  19.53   & 19.50 & 18.70 & 5.76 \\
C* & Ma97, Ma03 & 173   & 3.88& 0.535   &  1.287  &  37.51  & 0.37  &   0.37 & 0.00 & 0.00 & 0.00 & 0.23 & 0.28  &  14.75   & 19.88 & 18.98 & 5.97 \\
G & Non-Rotating & 0.00 & 3.53 & 0.493  &  1.287  &  36.36  & 0.40  &   0.37 & 0.00 & 0.00 & 0.00 & 0.22 & 0.28  &  13.11   & 19.84 & 18.37 & 5.74 \\
\multicolumn{18}{c}{} \\
\hline
\end{tabular}
}}

\end{table*}

\begin{figure}
   \centering
    \includegraphics[width=0.24\textwidth]{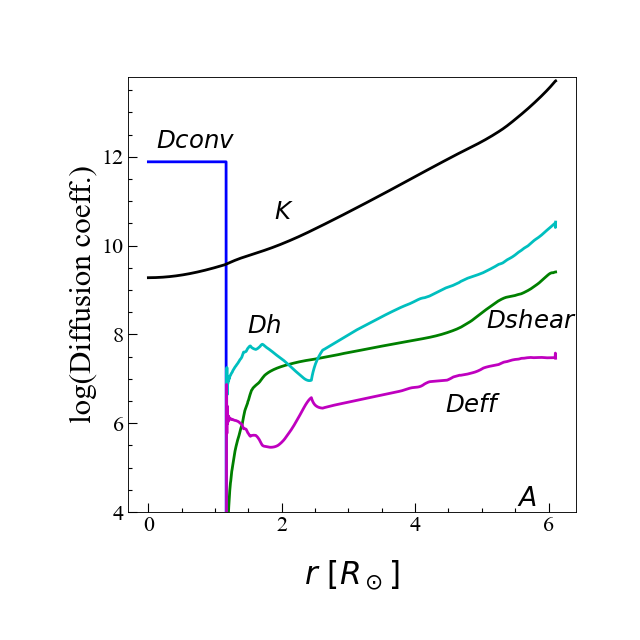}
    \includegraphics[width=0.24\textwidth]{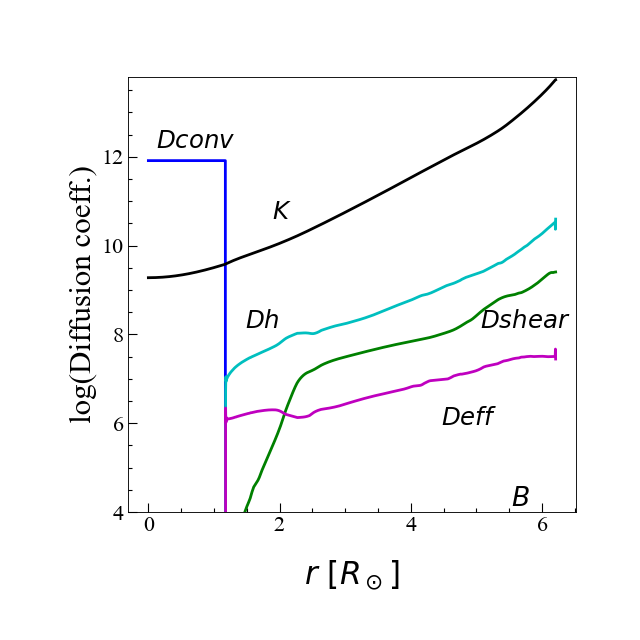}    
    \includegraphics[width=0.24\textwidth]{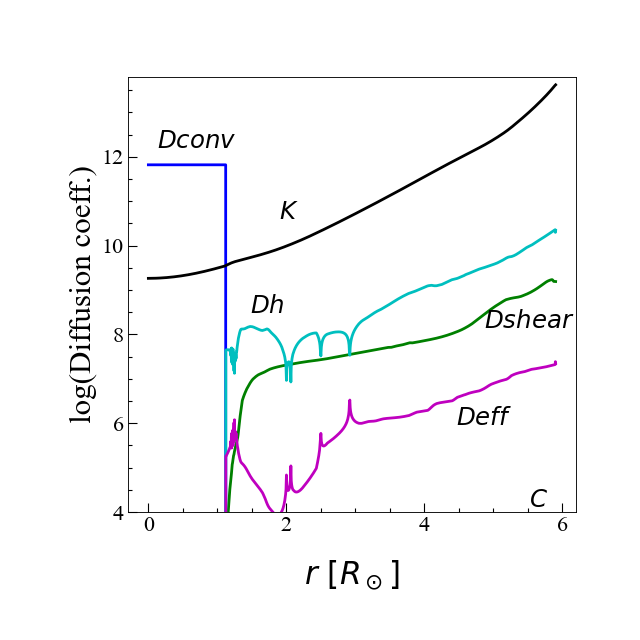}
    \includegraphics[width=0.24\textwidth]{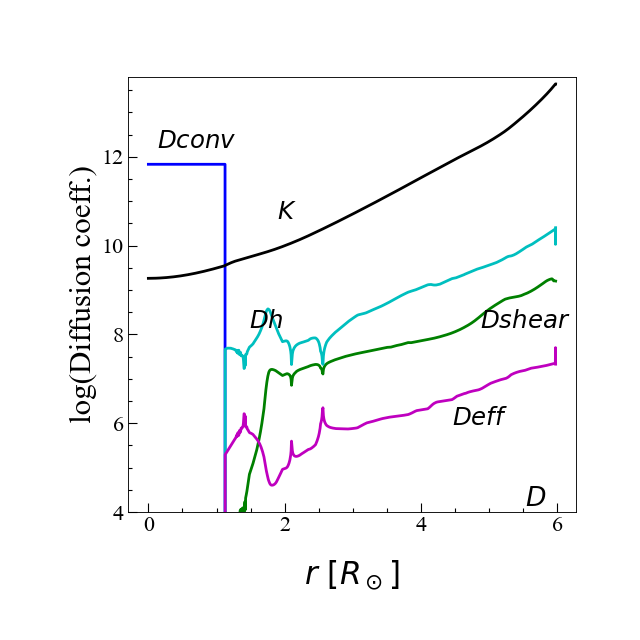}
   \includegraphics[width=0.24\textwidth]{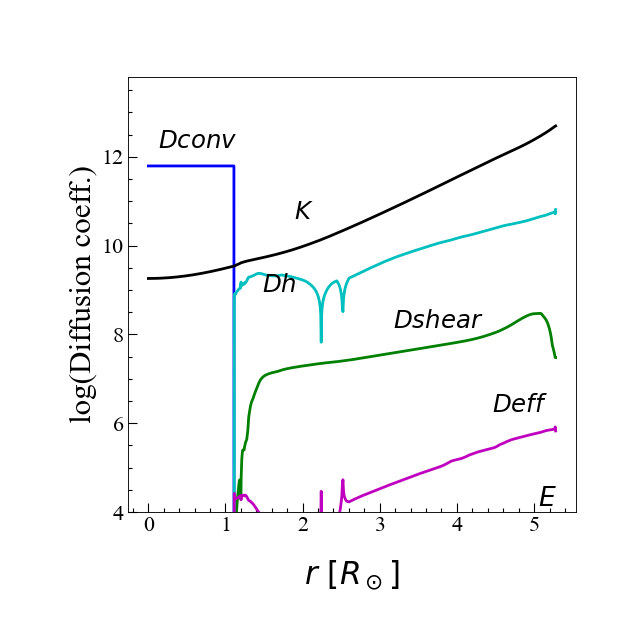}
    \includegraphics[width=0.24\textwidth]{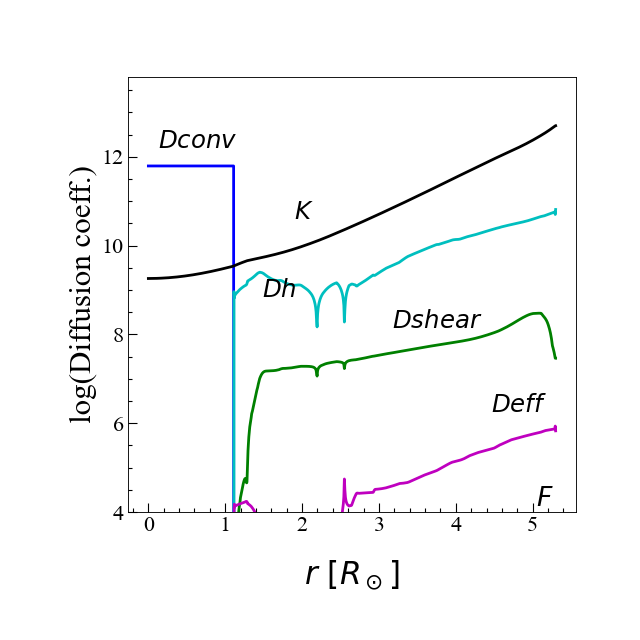}    
      \caption{Internal profiles of K the thermal diffusivity, D$_{\rm conv}$ the convective diffusion coefficient, D$_{\rm shear}$ the shear diffusion coefficient,D$_{\rm h}$ the horizontal turbulence coefficient and D$_{\rm eff}$ the effective diffusivity in 15 M$_\odot$ models at solar metallicity. Each panel is labeled with a letter corresponding to the first column of Table 1. The profile is taken when the central mass fraction of hydrogen X$_{\rm c}$ = 0.35.}
         \label{fig_15_DIFF}
   \end{figure}

\section{Diffusive/viscosity coefficients}\label{Sec:Coeff}

In this section, we discuss the expressions for the diffusive coefficients entering equations (1) and (2).
In the equation for the transport of the chemical species for non-magnetic models, $D_{\rm chem}$ is equal to the sum of two terms : $D_{\rm shear}$ and $D_{\rm eff}$. $D_{\rm shear}$ corresponds to the transport of both chemicals and angular momentum by the shear instability, while $D_{\rm eff}$ describes the transport of the chemical elements due to the combined action of meridional currents and horizontal turbulence \citep{Chaboyer1992}. In the equation for the transport of the angular momentum, $D_{\rm ang}$ is equal to $D_{\rm shear}$ for the non-magnetic models.

In magnetic models, $D_{\rm chem}$ is equal to the sum of three terms: $D_{\rm shear}$, $D_{\rm eff}$ and the direct transport of chemicals by the Tayler-Spruit dynamo $\eta_{\rm TS}$. In the equation for the transport of the angular momentum, $D_{\rm ang}$ is the sum of two terms:
$D_{\rm shear}$ and the viscosity associated to angular momentum by the Tayler-Spruit dynamo $\nu_{\rm TS}$. In magnetic models, the transport of angular momentum by the meridional currents is much smaller that the transport due to the magnetic coupling, therefore only the transport by the magnetic coupling is considered. The transport of the chemical species is dominated by
$D_{\rm eff}$ and therefore $\eta_{\rm TS}$ is neglected.

\subsection{The vertical shear diffusion, $D_{\rm shear}$}

The vertical shear diffusion coefficient $D_{\rm shear}$ accounts for the transport of chemical species and angular momentum in the radial direction, which is caused by turbulence triggered by differential rotation in the vertical direction. For this coefficient, we can find two different expressions in the literature; \citet{Maeder1997} and \citet{TZ1997}. 
Taken in the limit where the radiative diffusive coefficient $K$ is extremely large, and the shear instability has a negligible impact on energy transport, the expression of $D_{\rm shear}$ from \citet{Maeder1997} is:
  \begin{equation}
    D_\text{shear} =  f_\text{en} \frac{H_P}{g\delta}\frac{K}{\left[\frac{\varphi}{\delta}\nabla_\mu 
        + \left( \nabla_\text{ad} - \nabla_\text{rad} \right)\right]} 
    \left( \frac{9\pi}{32}\ \Omega\ \frac{\text{d} \ln \Omega}{\text{d} \ln r} \right)^2
  \end{equation}
where $f_\text{en}$ indicates the fraction of the available energy that can be used for the mixing. In the present models we have used  
$f_\text{en} = 1$., $H_P$ the pressure scale height, $g$ the gravity, $\varphi = \left( \frac{\partial\ln\rho}{\partial\ln\mu} \right)_{P,T}$, $K = \frac{4ac}{3\kappa}\frac{T^4\nabla_\text{ad}}{ \rho P \delta}$ is the radiative diffusion coefficient, $\delta=-\left( \frac{\partial\ln\rho}{\partial\ln T} \right)_{P,\mu}$, $\nabla_\mu=\frac{\text{d}\ln\mu}{\text{d}\ln P}$, 
$\nabla_{\rm ad}=\left(
\frac{\text{d}\ln T}{\text{d}\ln P}\right)_{\rm ad}$,
$\nabla_{\rm rad}=\left(
\frac{\text{d}\ln T}{\text{d}\ln P}\right)_{\rm rad}$. 
The expression of $D_{\rm shear}$ from \citet{TZ1997} is

  \begin{equation}
    D_\text{shear} =  f_\text{en} \frac{H_P}{g\delta}\frac{\left(K+D_\text{h}\right)}{\left[\frac{\varphi}{\delta}\nabla_\mu\left(1+\frac{K}{D_\text{h}}\right) + \left( \nabla_\text{ad} - \nabla_\text{rad} \right)\right]} \left( \frac{9\pi}{32}\ \Omega\ \frac{\text{d} \ln \Omega}{\text{d} \ln r} \right)^2,
  \end{equation}
where $D_h$ the horizontal turbulence diffusion coefficient.  

\subsection{The horizontal turbulence, $D_{\rm h}$}

The strong horizontal turbulence is triggered by shear along an isobar.
Differential rotation on an isobar can result from the action of meridional currents. Various expressions have been proposed by different authors, They are reminded below. They contain parameters whose values have been taken as proposed in the original publications. The interested reader may refer to these publications for the justification of these choices.

We can find three different expressions for $D_{\rm h}$ in the literature. \citet{Zahn1992} expresses
\begin{equation} \label{eqn_zahn_dh}
D_\text{h} =   r\ \left| 2\,V(r) - \alpha\,U(r) \right| 
\end{equation}
where $\alpha = \frac{1}{2} \frac{\text{d} \ln (r^2 {\Omega})}{\text{d} \ln r}$ and $V(r)$ is the horizontal component of the meridional circulation velocity\footnote {The original formula by \citet{Zahn1992} is multiplied by $\frac{1}{c_\text{h}}$, but the $c_h$ is then taken equal to 1}.
\citet{Maeder2003} expresses the equation as
\begin{equation}
D_\text{h} =  A\ r\ \left( r{\Omega}(r)\ V\ \left| 2V-\alpha U \right| \right)^{1/3} \\
\end{equation}
with $\alpha$ as in Eq.~\ref{eqn_zahn_dh}, and $A=0.002$.
\citet{Mathis2004} express $D_{\rm h}$ as
\begin{equation}
D_\text{h} =  \left( \frac{\beta}{10} \right)^{1/2} \left( r^2{\Omega} \right)^{1/2} 
\left( r \left| 2V-\alpha U \right| \right)^{1/2}
\end{equation}
with $\alpha$ as in Eq.~\ref{eqn_zahn_dh} and $\beta=1.5\cdot10^{-6}$. In all the above expressions, $\Omega$ is the latitude-averaged value of the angular velocity along an isobar. The variations of the angular velocity with respect to the latitude on a given isobar are expected to very small as a result of the action of the strong horizontal turbulence.

\subsection{The effective diffusion, $D_{\rm eff}$}

All prescriptions use the same effective mixing coefficient, $D_{\rm eff}$ for the chemical species given by \citet{Chaboyer1992, Zahn1992}:
\begin{equation} \label{eqn_deff}
  D_\text{eff} =  \frac{1}{30} \frac{\left| r\ U_2(r) \right|^2}{D_\text{h}}.
\end{equation}

As briefly reminded in Appendix \ref{Sec:ApA}, the factor $1/30$ results from an integration and is not a parameter whose value can be chosen.

Although the expression of $U$ shows a dependency on $D_{\rm h}$, this dependency remains small as long as $D_{\rm h}$ is much smaller than $K$.
This is the the case in all of our models. 
As indicated by Equation \ref{eqn_deff}, $D_{\rm h}$ is inversely proportional to $D_\text{eff}$. 
The horizontal turbulence limits any vertical movement similar to how a strong horizontal wind would bend the trajectory of smoke coming out of a chimney.

\subsection{The magnetic diffusivity, $\eta_{\rm TS}$}\label{Sec:CoeffMag}

The magnetic models used in this paper are based on Tayler-Spruit calibrated dynamo described in \citet{Eggenberger2022}. This is an approach built upon the theory of the Tayler-Spruit dynamo \citep{Spruit2002}, and calibrated in order to 
fit the internal distributions of angular velocity in subgiants and red giant stars as deduced from asteroseismic analysis.

The transport of the chemical elements by the magnetic instability is described by a magnetic diffusion coefficient, $\eta_{\rm TS}$. The magnetic diffusion coefficient is in absence of any instability the ohmic diffusion coefficient as given in \citet{Spitzer1956}. When the condition for the magnetic instability is realized, the values of $\eta_{\rm TS}$ and of the Alfven frequency can be obtained from two conditions valid just at the limit when the instability triggers \citep[see the discussion in][]{Eggenberger2022}.

This magnetic diffusion is not very important compared to the transport of the chemical species by $D_{\rm eff}$. It is is neglected in our models. However, this quantity needs to be computed since it is used in the expression for the magnetic viscosity (see below).

\subsection{The magnetic viscosity, $\nu_{\rm TS}$}

The transport of the angular momentum is described by a magnetic viscosity parameter $\nu_{\rm TS}$. The viscosity is then obtained using the general formula given by Eq.~(8) of \citet{Eggenberger2022}

 \begin{equation}
 \nu_{\rm TS} =
  \; \frac{\Omega \; r^2}{q} \;
 \left(  C_{\rm T} \; q \; \frac{\Omega}{N_{\rm eff}}  \right)^{3/n} \; 
\left(\frac{\Omega}{N_{\rm eff}}\right) \; ,
\end{equation}
with $n=1$, $C_{\rm T}=216$ and  $N_{\rm eff}^{2}={\eta \over K} N_{T}^{2} +N_{\mu}^{2}$,
$N_T^2=g\delta/H_p (\nabla_{\rm ad}-\nabla_{\rm rad})$ and $N_\mu^2=g/H_p\nabla_\mu$. The value of $C_{\rm T}$ has been chosen so that the models can reproduce the core rotation rates of red giants determined by asteroseismology by \citet{Gehan2018}. In the original version of the Tayler-Spruit dynamo  $C_{\rm T}=1$ \citep{Spruit2002}. The version with $C_{\rm T}=216$ has been called the calibrated Tayler-Spruit dynamo by \citet{Eggenberger2022}. The calibrated TS dynamo provides an evolution of the core rotation similar to that given by the \citet{Fuller2019} approach.
This magnetic angular momentum transport is only accounted for when the shear parameter $q= \left| \frac{\partial \ln \Omega}{\partial \ln r} \right| $ is larger than a minimum value $q_{\rm min}$ given by Eq.~(12) of \citet{Eggenberger2022}:

 \begin{equation}
q_{\rm min} = C_{\rm T}^{-1} \left(\frac{N_{\rm eff}}{\Omega}\right)^{(n+2)/2} \left(\frac{\eta}{r^2 \Omega}\right)^{n/4} \; .
\label{qmin_generale}
\end{equation}

\subsection{Parameters of stellar models}

Additional parameters of the stellar models are chosen as described in \citet{Ekstrom2012a}. In particular, the models are computed with a moderate step overshoot ($\alpha_{\rm ov} = 0.1$), use the same mass loss rates and no surface magnetic fields are accounted for. The convective zones rotate as solid bodies and are chemically homogeneous. More details on the models can be found in Appendix\ref{Sec:ApB}.

We have chosen to focus on two initial masses representative of the evolution of massive stars. The first is a 15 M$_\odot$ model that illustrates the case of single stars evolving into a red supergiant stage after the Main-Sequence and exploding as a type IIP supernova at the end of their evolution. The second is the case of a single 60 M$_\odot$ model evolving into a Luminous Blue Variable phase after the Main-Sequence and evolving into a Wolf-Rayet phase. This is representative of the upper mass range of single massive stars whose evolution at solar metallicity is mainly affected by mass loss. All the models are computed using the fully advective-diffusive approach (according to Equation~\ref{eqn_ang_mom}) during the MS phase and using a purely diffusive approach during the core helium burning phase. 

In the present work, we did not impose values to uncertain parameters involved in the expressions for the diffusion coefficients to enforce the model to reproduce a given observational features as commonly done when grids of rotating models are computed.
Instead we considered for all the models the same physical assumption for setting the parameter $f_{\rm en}$ in $D_{\rm shear}$ and opted otherwise the values of the parameters $A$ or $\beta$ in the different expression for $D_{\rm h}$ suggested by the original publications. For setting $f_{\rm en}$, we assumed for all the models that the whole excess energy available in the differential rotation (shear energy) can be used to drive the transport and that the critical Richardson number is taken equal to 1/4 (for a detailed discussion on the physics of $f_{\rm en}$ and Richardson, please refer to section \ref{Sec:ApA} and \ref{Sec:ApB} respectively). These two assumptions have no reason to vary when we use different prescriptions since all of them are based on similar physics. 

Thus what we are doing here is to compare the impacts of different physical approaches without any fine tuning to reproduce an observed feature. The ouputs of the models are however comparable since at a more fundamental level they are based on common physical assumptions.

Table~\ref{tab:1} shows all the models discussed in the present paper. 
The first column gives the model label. The prescription used is given in column 2: Ma97, TZ97 means that the expression for the shear diffusion is given by respectively \citet{Maeder1997}, \citet{TZ1997}.
Za92, Ma03, and MZ04 indicates that the expression of the horizontal diffusion coefficient is given by respectively by
\citet{Zahn1992}, \citet{Maeder2003} and \citet{Mathis2004}.
The time-averaged equatorial 
velocity during the MS phase is given in column 3, the MS lifetime is given in column 4, the surface helium abundance in mass fraction  at the end of the MS phase is given in column 5.
Column 6 presents the N/H ratio obtained at the surface, at the end of the MS phase,  and normalized to the initial N/H value. 
The actual mass of the star at the end of the core H-burning phase is indicated in column 7. The core He-burning lifetime, and the durations of the core He-burning phase spent in the red ($\log T_{\rm eff} < 3.68$), in the blue ($\log T_{\rm eff} > 3.87$) and in the yellow part ($3.68 < \log T_{\rm eff} < 3.87$) of the HR diagram are given in columns 8, 9, 10, and 11 respectively. The ratio of the time spent in the blue to that spent in the red is shown in column 12. The mass fraction of helium at the surface is given in column 13, the actual mass at the end of the core He-burning lifetime in column 14. The masses of the helium cores, of the carbon-oxygen cores () and of the remnants are given in columns 15, 16 and 17 and are obtained using the formulation from \cite{Maeder1992}. The masses of the helium and CO cores are obtained finding the first shell scanning the star from the surface to the interior, where the mass fraction of helium, respectively the sum of the mass fraction of carbon and oxygen is larger than 0.75).
Model C* is computed using the same prescription as model C but additionally includes the effects of magnetic fields as described in section \ref{Sec:CoeffMag}

Some differences between models of the same initial mass computed with different prescriptions for the rotational mixing are small, small in the sense that some change in the numerics (time or space resolution) can produce similar changes and small also in the sense that they cannot be discriminate by observations. This concerns mainly differences due to the different prescriptions on the HR track during the MS phase.
While the reader has to keep in mind this when looking at the present results, we would however stress that the technics governing the choice of the time and space resolution (briefly described in Appendix B) has been kept exactly the same for all the computations, allowing thus to give hopefully a correct idea of the relative changes brought by the different prescriptions.

\section{The results of non-magnetic 15 M$_\odot$ models}\label{Sec:Results_nonmag}   

\begin{figure*}
   \centering
    \includegraphics[width=0.33\textwidth]{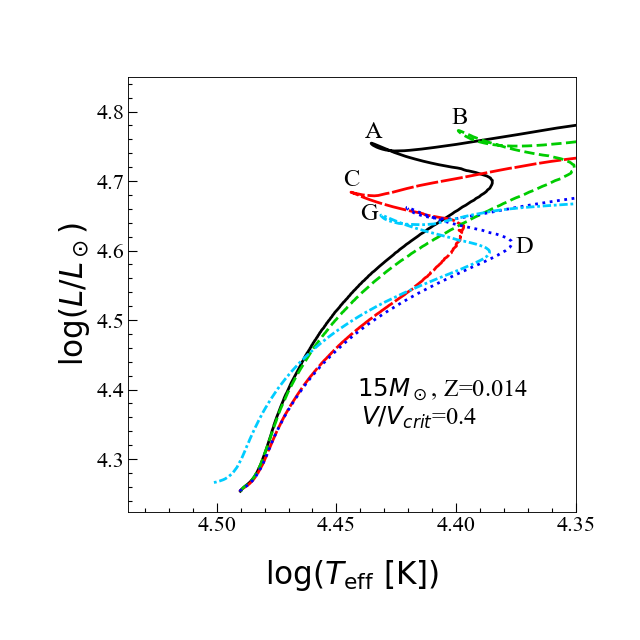}
    \includegraphics[width=0.33\textwidth]{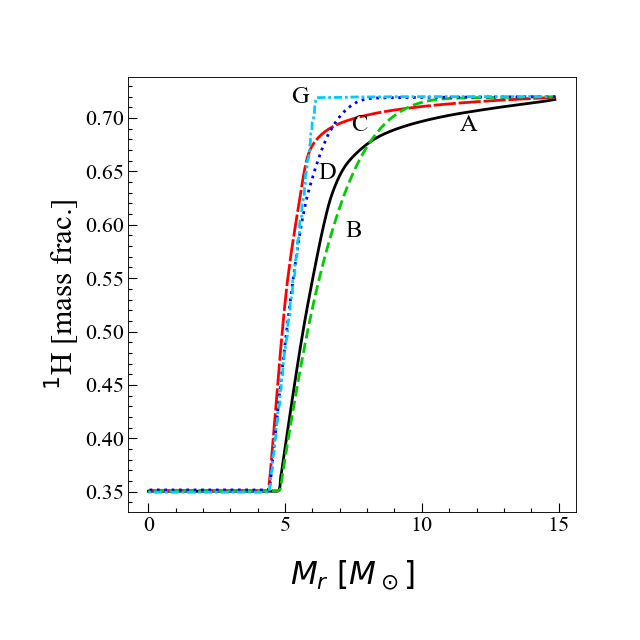}
    \includegraphics[width=0.33\textwidth]{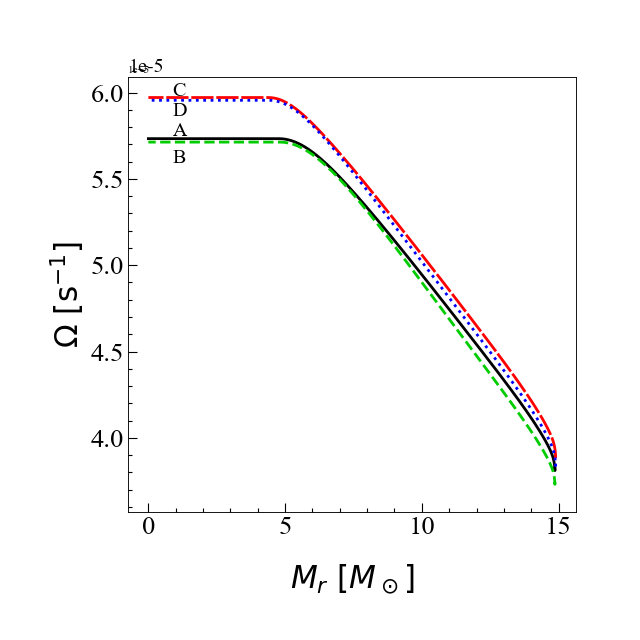}    
\caption{{\it Left panel}: Evolutionary tracks during the MS phase in the Hertzsprung-Russell diagram of the 15 M$_\odot$ at $Z$=0.014 and with $V_{\rm ini}/V_{\rm crit}=0.4$. The letters `A',`B',`C', `D' and `G' correspond to the models described in Table \ref{tab:1}.
{\it Centre panel}: variation of the mass fraction of hydrogen as a function of the Lagrangian mass 
coordinate when central mass fraction of hydrogen is 0.35 (Xc=0.35). 
{\it Right panel}: Evolution of angular velocity versus the internal mass coordinate at Xc=0.35 through hydrogen burning.}
         \label{fig_Z14_MS}
   \end{figure*}

\subsection{The diffusion coefficients}

Figure~\ref{fig_15_DIFF} illustrates the variation of diffusion coefficients with radius when the central mass fraction of hydrogen is 0.35.

Among the qualitative features that are similar in all six models we have:
\begin{itemize}
\item In the convective core, the diffusion coefficient is large enough for homogenizing the chemical composition. Here it is
of the order of 10$^{12}$ 
cm$^2$ s$^{-1}$. This is due to the physics of convection and not of rotation.
\item The thermal diffusion coefficient $K$ increases outwards and is well above the diffusion coefficients describing the rotational instabilities. Thus, the thermal diffusion timescale is much shorter than the timescales of the rotational instabilities, indicating that the rotational instabilities do not significantly contribute to the transport of energy.
\item $D_{\rm h}$ serves as the primary rotational diffusion coefficient in the majority of the radiative region, and it is strong enough to achieve shellular rotation.
\end{itemize}

Rotational mixing can facilitate the transfer of an element from the core to the surface. During the MS phase, an element that is transported from the core to the surface passes through four zones. The first zone is the convective core where the transport of the chemical species is very efficient.

The second zone is located immediately above the core, and it is characterized by a gradient in the mean molecular weight. 
The chemical gradient comes from the fact that the nuclear timescale is shorter than the mixing timescale in the the whole radiative envelope.
The gradient is steeper in the layers directly surrounding the core and then it becomes smoother when progressing outwards (see the middle panel of Fig.~\ref{fig_Z14_MS}). 
This comes from the fact that during the MS phase, the convective core decreases in mass. The outer portion of this zone is therefore representative of the initial impact of the nuclear reactions in the core.
In the region with a steep $\mu-$gradient, the vertical shear diffusion coefficient is strongly reduced by the $\mu-$gradients and the main diffusion coefficient for transporting the chemical species is 
$D_{\rm eff}$. The $D_{\rm eff}$ is smaller when the value of $D_{\rm h}$ is larger. Thus the choice of $D_{\rm h}$ has a significant impact on the
way diffusion of the chemical species occurs in the second region.
The third zone is the region where  
${\varphi \over \delta}\nabla_\mu$ is significantly larger than the difference $\nabla_\text{ad} - \nabla_\text{rad}$, the ratio
$D_\text{shear}\text{(M97)}/D_\text{shear}\text{(TZ97)} \sim K/D_\text{h}$.
Since $D_\text{h}$ is less than $K$, one has $D_\text{shear}\text{(M97)} > D_\text{shear}\text{(TZ97)}$. The mixing in this region therefore depends on the vertical shear diffusion coefficient. 
Finally, zone 4 covers all the outer layers above the zone 3. Here the mixing of the elements is also governed by $D_{\rm shear}$, but the value of $D_{\rm shear}$ is the same whether \citet{Maeder1987} or \citet{TZ1997} is used. Indeed, in zones with no $\mu$-gradients, $D_\text{shear}\text{(M97)}/D_\text{shear}\text{(TZ97)} \sim K/(K+D_\text{h}) \sim 1$, since $D_{\rm h}$ is smaller than K.

\subsubsection{The vertical shear diffusion coefficient D$_{\rm shear}$}

In the first two upper panels of Fig.~\ref{fig_15_DIFF}, we compare the impact of transitioning from the expression of \citet{Maeder1997} to that of \citet{TZ1997} for D$_{\rm shear}$, while keeping the same expression of \citet{Zahn1992} for D$_{\rm h}$. We see that even if the same expression for $D_{\rm h}$ is used, the profile of $D_{\rm h}$ inside the two models is not equivalent (see models A and B). This is due to the fact that changing D$_{\rm shear}$ modifies the chemical structure of the model and hence its structure at a given evolutionary stage. The same is true for D$_{\rm eff}$. We note that when the D$_{\rm shear}$ of \citet{Maeder1997} is used, the zone in which $D_{\rm eff}$ dominates is significantly reduced. 
This is due to the impact of the stronger value given by the expression from 
\citet{Maeder1997}. Note also that the use of the D$_{\rm shear}$ by \citet{Maeder2003} reduces the build up of $\mu-$gradients compared to \citet{TZ1997} as there are only weak $\mu-$gradients at the beginning of the evolution.

In the outer regions of the radiative zone, we can see that both expression gives similar values as expected. 
On the whole, models using the D$_{\rm shear}$ of \citet{Maeder1997} will show a stronger mixing at the end of the MS phase than the model using the expression by \citet{TZ1997}. The impact on the tracks in the Hertzsprung-Russel (HR) diagram will be discussed in Section \ref{Sec:Tracks&Life15}.

\subsubsection{The horizontal shear diffusion coefficient D$_{\rm h}$}

The middle panels of Fig.~\ref{fig_15_DIFF} show the diffusion coefficients in models at the middle of the MS phase using the expression of $D_{\rm h}$ given by \citet{Maeder2003}.
In such models, $D_{\rm eff}$ is decreased due to the larger value of $D_{\rm h}$.
The model using D$_{\rm shear}$ from \citet{Maeder1997} is more mixed than the model using the D$_{\rm shear}$ from \citet{TZ1997}. This is a similar qualitative behavior as observed when the
$D_{\rm h}$ of \citet{Zahn1992} is used. 

The lower panels of Fig.~\ref{fig_15_DIFF} display the outcomes for diffusion coefficients with the $D_{\rm h}$ \citet{Mathis2004}, which yields greater magnitudes than in the previously discussed models.
Consequently, the values of $D_{\rm eff}$ are considerably decreased as $D_{\rm h}$ approaches $K$. As a result, the difference between the outcomes obtained using the two distinct expressions of $D_{\rm shear}$ is reduced.

The total diffusion coefficient for the chemical species is the sum of
$D_{\rm eff}$ and $D_{\rm shear}$. 
$D_{\rm eff}$ plays a crucial role in determining the events at the edge of the convective core, and adjustments in the mixing efficiency in this area can influence the size of the convective core. We will discuss this point and the impact on the surface abundances in more detail below. It is worth noting that even if $D_{\rm eff}$ were to be zero, surface enrichment would still occur. This is because during the MS phase, the convective core mass recedes over time, creating a region with a $\mu$-gradient in which $D_{\rm shear}$ dominates the mixing process, as explained earlier.

\subsection{Tracks and lifetimes}\label{Sec:Tracks&Life15}

\begin{figure}
   \centering
    \includegraphics[width=0.24\textwidth]{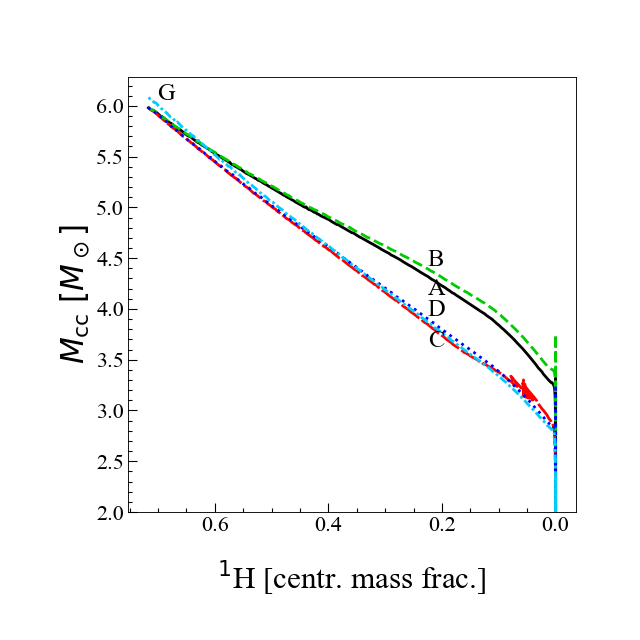}
    \includegraphics[width=0.24\textwidth]{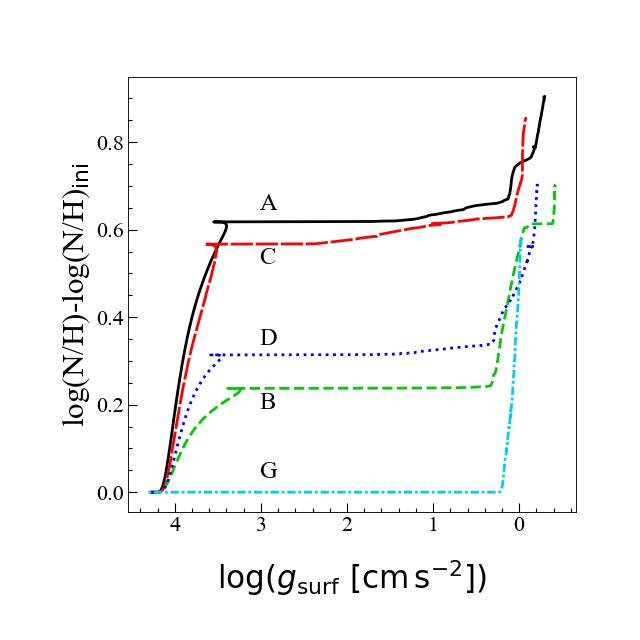}
      \caption{{\it Left panel}: Mass of the convective core in solar mass as a function of central mass fraction of hydrogen X$_c$ in different 15 M$_\odot$ models. {\it Right panel}: Evolution of the surface abundance ratio N/H (normalised to its initial value and in logarithmic scale) as a function of log g$_{surf}$ for the same models as those presented in the left panel. In both panels, 
      the letters have the same meaning as in Fig.~\ref{fig_Z14_MS}.}
         \label{fig_Z14_CCNH}
   \end{figure}

In the left panel of Fig.~\ref{fig_Z14_MS}, the evolutionary tracks for rotating 15~M$_\odot$ models with different prescriptions for rotation are compared. The non-rotating track is also plotted (cyan dashed-dotted line labeled G). 
As is well known from previous works \citep[see e.g.][]{Faulkner1968, Kip1970, Endal1976, MM1997}, the ZAMS position is shifted to lower values of effective temperature and luminosity as a result of the hydrostatic effect of rotation (note that at ZAMS the star is still chemicall homogeneous thus mixing effects are not yet present.). The hydrostatic effect on the ZAMS, as expected, do not depend on the prescriptions used for the diffusion coefficients.

\subsubsection{Impact of changing D$_{\rm shear}$}

Comparing models A and B in the left panel of Fig.~\ref{fig_Z14_MS}, we can see the changes caused by switching the expression of D$_{\rm shear}$ from \citet{Maeder1997} to that of \citet{TZ1997}. 
Model A is bluer and more luminous than model B. This is a result of A being more mixed than model B, due to the larger value of D$_{\rm shear}$ in zone 3 in A.

The middle of Fig.~\ref{fig_Z14_MS} indicates that there is more mixing in Model A than Model B as evidenced by the lower hydrogen mass fraction below the surface. Furthermore, the nitrogen surface enrichment in Model A is significantly more pronounced than in Model B, as can be observed from the right panel of Fig.~\ref{fig_Z14_CCNH}. This can be attributed to the increased chemical element transport in Model A, which causes a greater inward diffusion of hydrogen in the radiative zone. Therefore, for comparable $D_{\rm eff}$ values, more hydrogen tends to accumulate in the layers surrounding the convective core in Model A than in Model B, leading to a steeper gradient in the former. This demonstrates that a model with more significant mixing in one specific region may imply steeper gradients in others.

The position of the Main-Sequence band in the HR diagram is influenced not only by the chosen overshooting parameter but also by the rotation physics employed. Different rotation models can shift this limit, highlighting the need to consider the rotational mixing physics when analyzing the Main-Sequence band. Although the convective core masses in Models A and B display comparable evolution as illustrated in the left panel of Fig.~\ref{fig_Z14_CCNH}, their TAMS positions are not identical (see the left panel of Fig.~\ref{fig_Z14_MS}). Therefore, when calibrating models for the overshooting parameter using the TAMS position in the HR diagram, it is essential to keep in mind the potential influence of the rotation physics on this position \citep[see the discussion in][]{Martinet2022}.

\subsubsection{Impact of changing D$_{\rm h}$}
We will now compare Models C and D which have the same $D_{\rm shear}$. These models exhibit a generally larger $D_{\rm h}$ value than in Models A and B, while featuring a smaller $D_{\rm eff}$ value.
This produces smaller convective cores as seen in the left panel of Fig. \ref{fig_Z14_CCNH}. Models are less mixed chemically and this makes the tracks less luminous than Models A and B. The difference between the tracks C and D are otherwise qualitatively similar than between Models A and B.  
We see that the Model C, despite having a significant lower value of $D_{\rm eff}$ just above the core still shows strong surface enrichment (see the right panel of Fig.~\ref{fig_Z14_CCNH}).
This is a consequence of the fact that the convective core recedes in mass as explained above.
In Fig.~\ref{fig_Z14_CDEF_1} in the appendix, the tracks in the HR diagram for the Models E and F are compared to the Models A, B, C and D. Model E follows a very similar path as Model C. Model F presents similar characteristic with Model D although being slightly less luminous and having a bluer turn off point.

\subsubsection{Impact on main sequence lifetime}

The MS lifetimes for the 15 M$_\odot$ stellar models are indicated in the fourth column of Table~\ref{tab:1}. For all the models, the duration of the core hydrogen burning phase is increased by rotation by 1 - 22\% compared to the non-rotating model. The rotating models using the value of $D_{\rm h}$ by \citet{Maeder2003} and the D$_{\rm shear}$ by \citet{TZ1997} (models D and F) have the smallest increase in MS lifetime (between 1 and 6\%).

\subsection{Internal and surface rotations, surface abundances}

In the non-magnetic models, the main process for transporting angular momentum is the meridional circulation. In all our models, we used the same equation for computing the meridional currents. Only the expression of $D_{\rm h}$ enters into the equation giving $U_2$, appearing as a multiplying factor (1+$D_{\rm h}$/K). However, since $D_{\rm h}$ is in general at least an order of magnitude below $K$ in our models, the change of $D_{\rm h}$ has only a limited effect. Hence, this effect is unlikely to have a significant impact on the transport of angular momentum. The right panel of Fig.~\ref{fig_Z14_MS} shows the internal rotation of the different models at the middle of the MS phase to be quite similar. Models computed with a higher value of $D_{\rm h}$ have slightly faster rotating cores.

The time-averaged surface rotational velocity during the MS phase is indicated in the third column of Table~\ref{tab:1}. The median value is 188 km s$^{-1}$. Variations of $\pm$8\% are obtained depending on the expressions used and thus, the impact on the surface rotation rates remains modest. 
Changing the diffusion coefficient alters the chemical structure and thus impacts the stellar radius and mass loss by stellar winds, which impacts the surface rotational velocity.
However, the masses at the end of the MS phase among the different 15 M$_\odot$ models exhibit minimal variation, with differences of approximately $\pm$0.4\%, as shown in column 4 of Table~\ref{tab:1}\footnote{Note that uncertainties in the mass loss rates produces larger differences \citep{Renzo2017}}.
This explains why the differences of surface velocities remain very modest.

In contrast to the rotation rates, the surface enrichment is very sensitive to the choice of the diffusion coefficients (right panel of Fig.~\ref{fig_Z14_CCNH}). The choice of $D_{\rm shear}$ has a greater impact on surface enrichment compared to $D_{\rm eff}$, which governs exchanges between the convective core and the base of the radiative envelope. Models using the expression for $D_{\rm shear}$ by \citet{Maeder1997} exhibit much stronger enrichment at the end of the MS phase than those using the expression by \citet{TZ1997}, regardless of the chosen $D_{\rm h}$.

\subsection{He-burning lifetimes and blue-red supergiant ratios}

\begin{figure}
   \centering
    \includegraphics[width=0.24\textwidth]{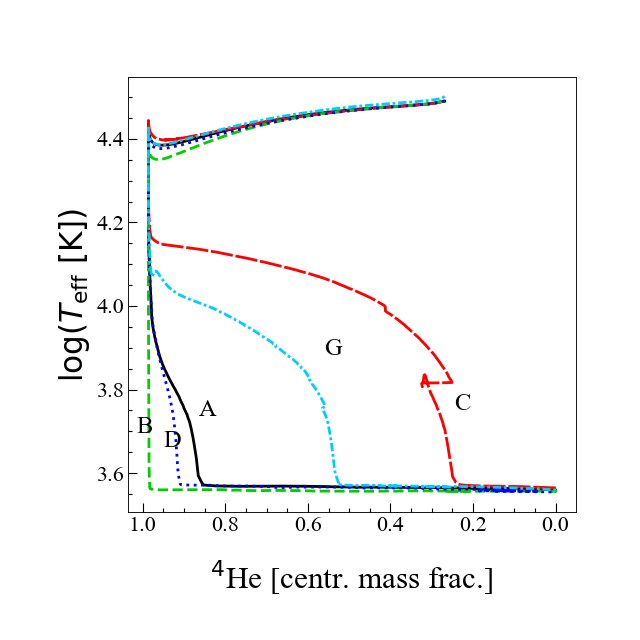}
    \includegraphics[width=0.24\textwidth]{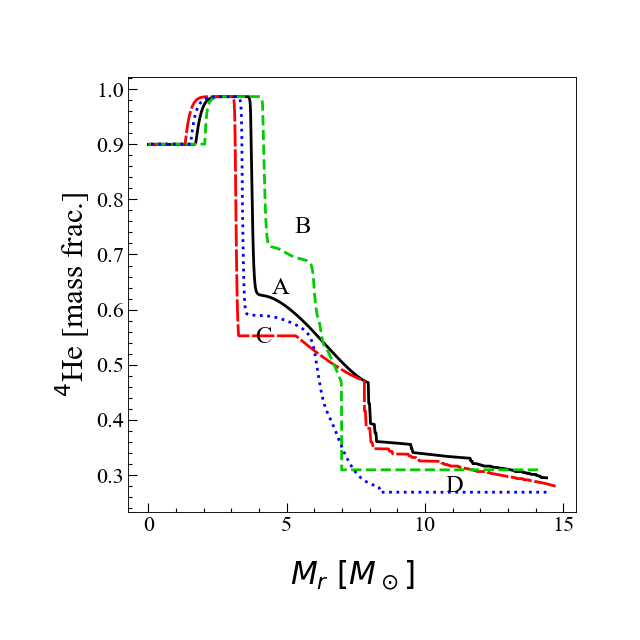}
      \caption{{\it Left panel}: Evolution of the effective temperature as a function of the central helium abundance for different 15M$_\odot$ models.{\it Right panel}: Helium profile versus the mass in different 15M$_\odot$ models when the central Helium mass fraction is 0.90. In both panels, 
      the letters have the same meaning as in Fig.~\ref{fig_Z14_MS}.}
         \label{fig_Omega_H1He4c_15_14T}
   \end{figure}

The core He-burning lifetimes are indicated in column 8 of Table~\ref{tab:1}. They range from 8 to 15\% of the MS lifetime. Larger values of $D_{\rm h}$ are associated with longer core He-burning phase and larger fractions of $\tau_{\rm He}/\tau_{\rm H}$.

The evolution of the effective temperature as a function of the mass fraction of helium at the centre of the star for the 15 M$_\odot$ models are shown in the left panel of Fig.~\ref{fig_Omega_H1He4c_15_14T}. Only Model C and, to a lesser extent, the non-rotating Model G show a different behaviour to the other models. These two models spend a significant fraction of the core He-burning phase with an effective temperature above 6000K ($\log T_{\rm eff}> 3.8$). Models evolving to the red supergiant stage directly do so on a Kelvin-Helmholtz timescale, while those staying longer in the blue region follow a nuclear-burning timescale. This affects the blue to red supergiant ratio and case B mass transfer in close binary evolution \citep[see][for the definition of a case B mass transfer]{KW1994, vanden1968}.

When the mass fraction of helium at the center is 0.9, the two models exhibit the lowest helium abundance above the H-burning shell. In models with a higher helium abundance above the He-core, a red position during the core He-burning phase is favored. This is in line with what was found in \citet{MM2001, Walm2015} and also in the numerical experiments by \citet{Eoin2022}.
It is interesting to note that the degree of mixing in the star as a whole is very different in models C and G. Despite this, they both favour a long duration in a blue supergiant phase during the first crossing of the HR gap.
This indicates that the time spent in different effective temperature ranges when the star crosses the HR gap depends on changes in the distribution of the chemical elements inside the star in and near the H-burning shell. This is in line with previous works as those by \citet{Schoot2019, Klencki2022}.

\subsection{Difference of the structure at the end of the core He-burning phase}

\begin{figure}
   \centering
    \includegraphics[width=0.24\textwidth]{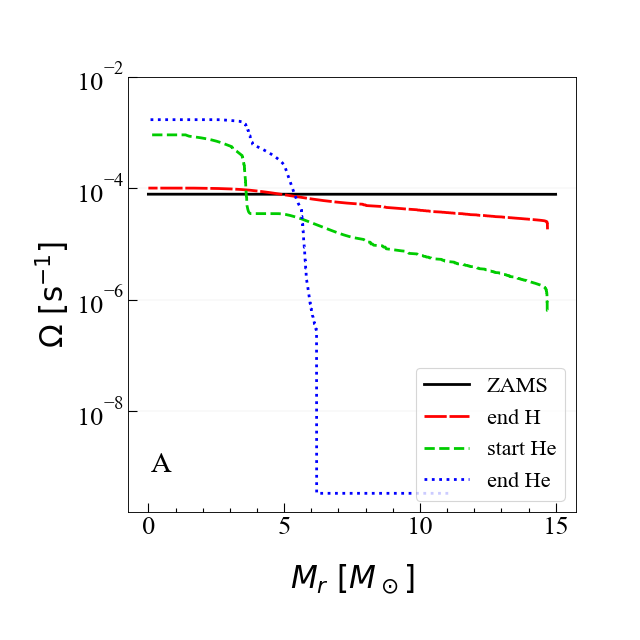}
    \includegraphics[width=0.24\textwidth]{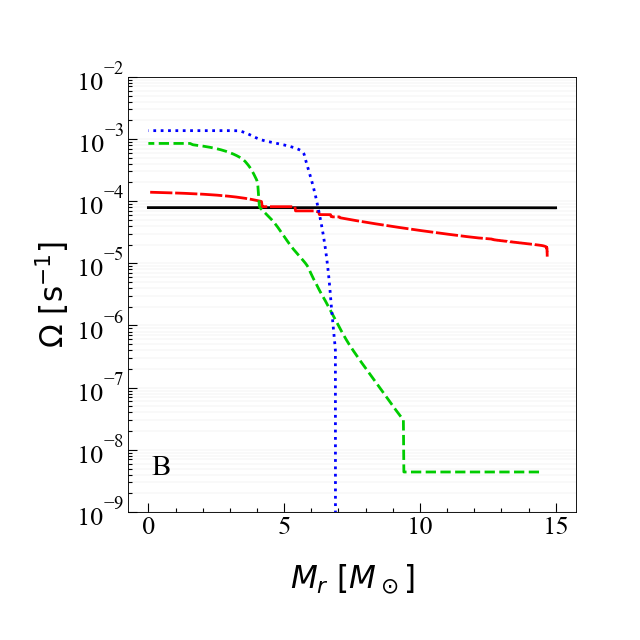}
    \includegraphics[width=0.24\textwidth]{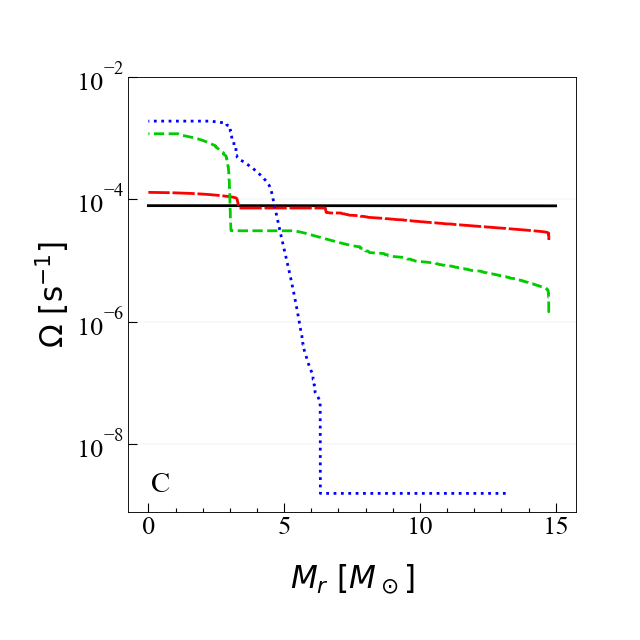}  
    \includegraphics[width=0.24\textwidth]{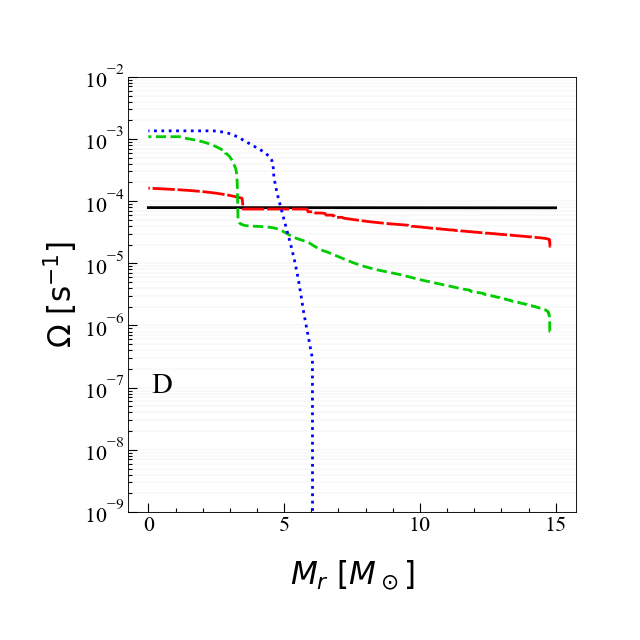}
    \includegraphics[width=0.24\textwidth]{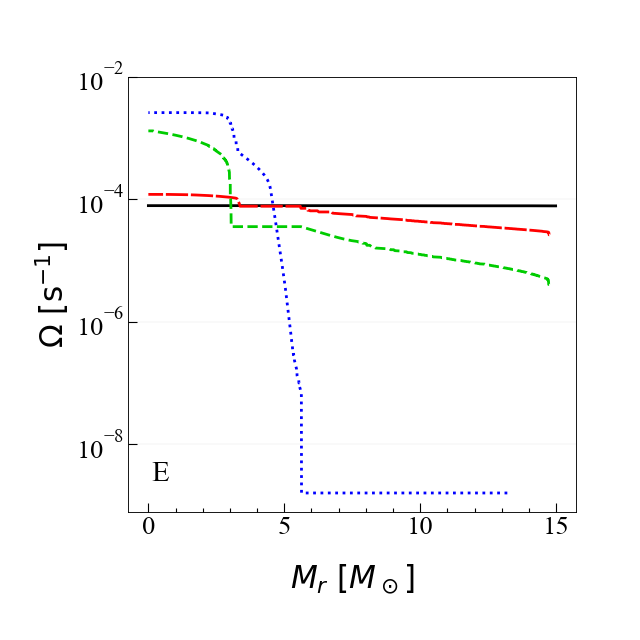}
    \includegraphics[width=0.24\textwidth]{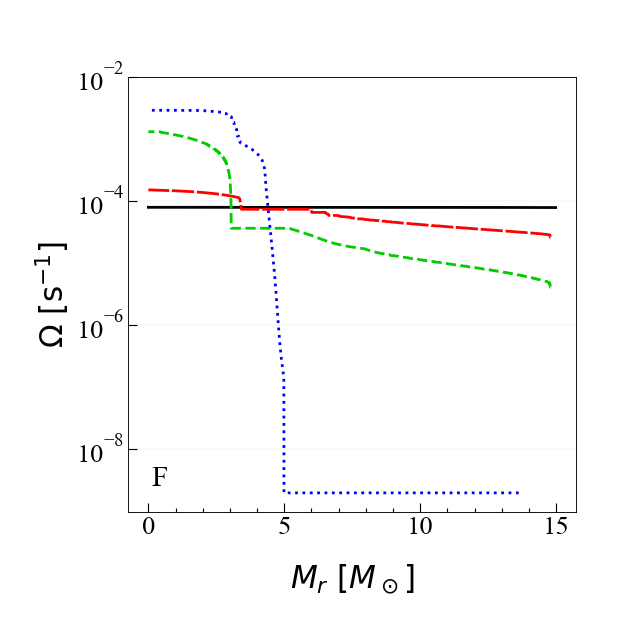}
      \caption{Angular velocity profiles as a function of mass for  different 15 M$_\odot$ models at various evolutionary stages.
      The letters have the same meaning as in Fig.~\ref{fig_15_DIFF}.}
         \label{fig_Omega_jr_15_A}
   \end{figure} 

 \begin{figure*}
   \centering
    \includegraphics[width=0.24\textwidth]{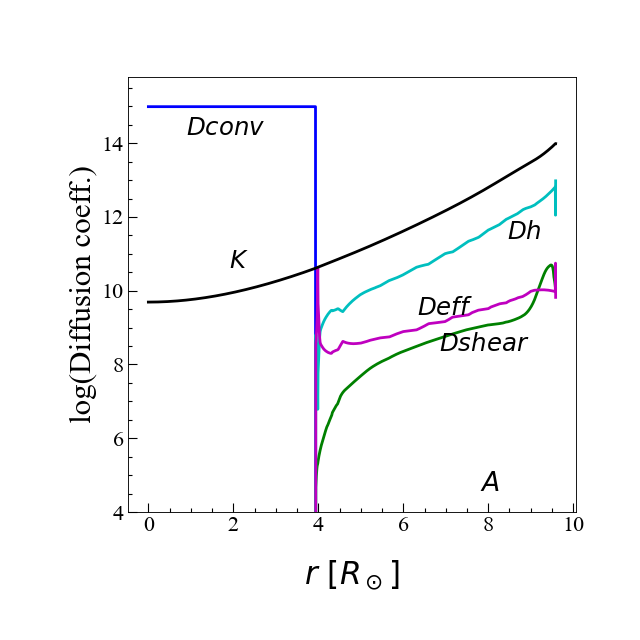}
    \includegraphics[width=0.24\textwidth]{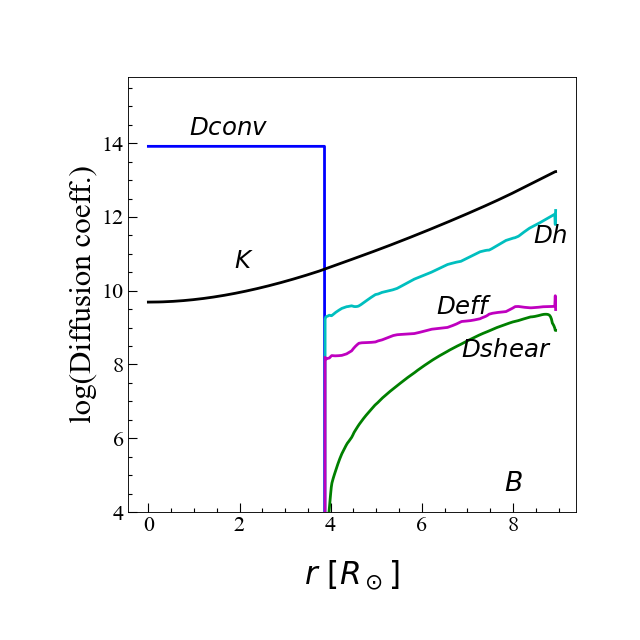}    \includegraphics[width=0.24\textwidth]{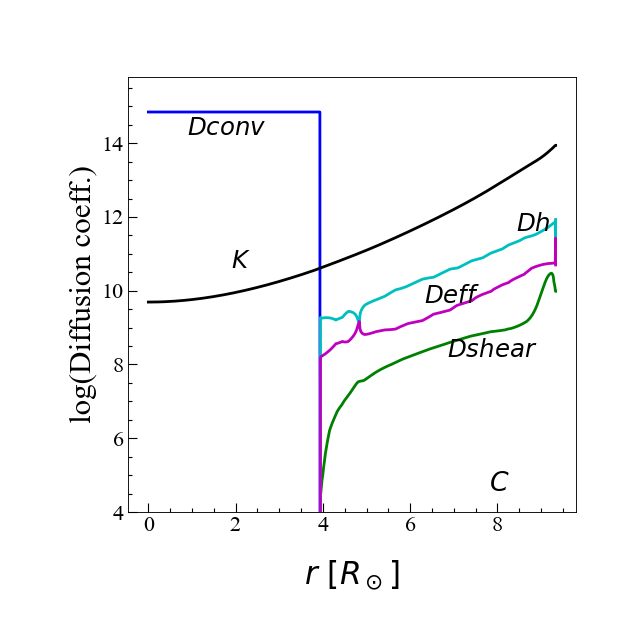}
    \includegraphics[width=0.24\textwidth]{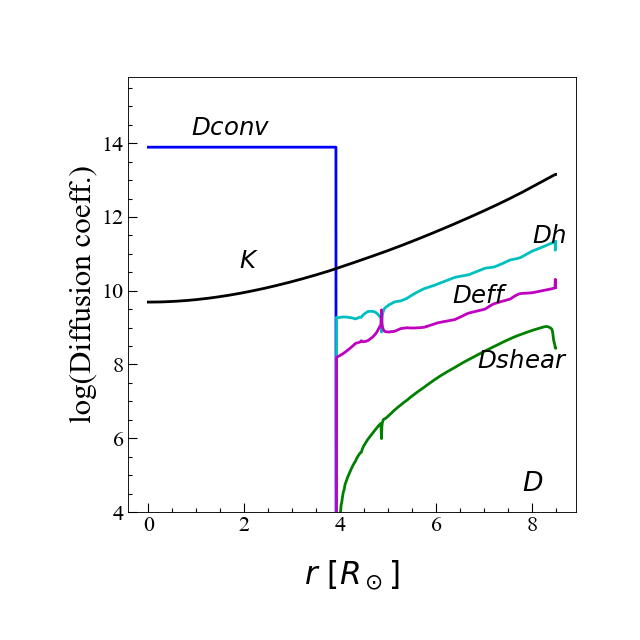}
      \caption{Variation as a function of the radius of
      K the thermal diffusivity, D$_{\rm conv}$ the convective diffusion coefficient, D$_{\rm shear}$ the shear diffusion coefficient,D$_{\rm h}$ the horizontal turbulence coefficient and D$_{\rm eff}$ the effective diffusivity in different  60 M$_\odot$ models at solar metallicity.  The letters correspond to models computed with the prescription given in the second column of Table~\ref{tab:1}. The profiles are taken when the central mass fraction of hydrogen X$_{\rm c}$ = 0.35.}
         \label{fig_D_60_14}
   \end{figure*}   

 \begin{figure*}
   \centering
    \includegraphics[width=0.33\textwidth]{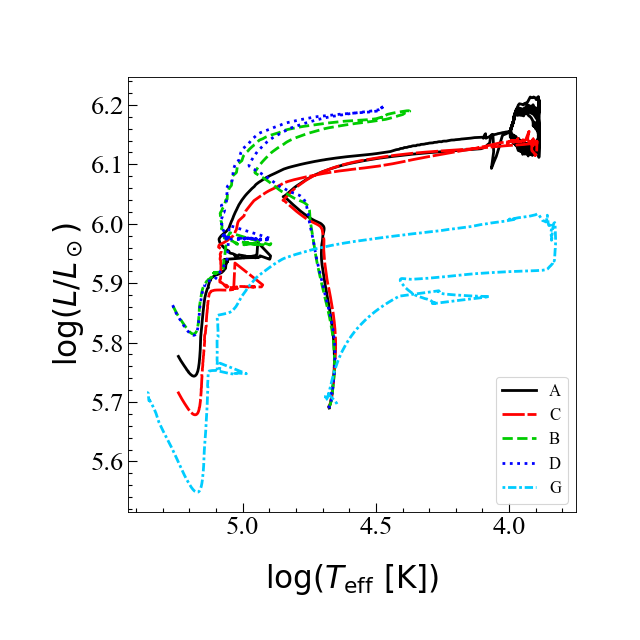}
    \includegraphics[width=0.33\textwidth]{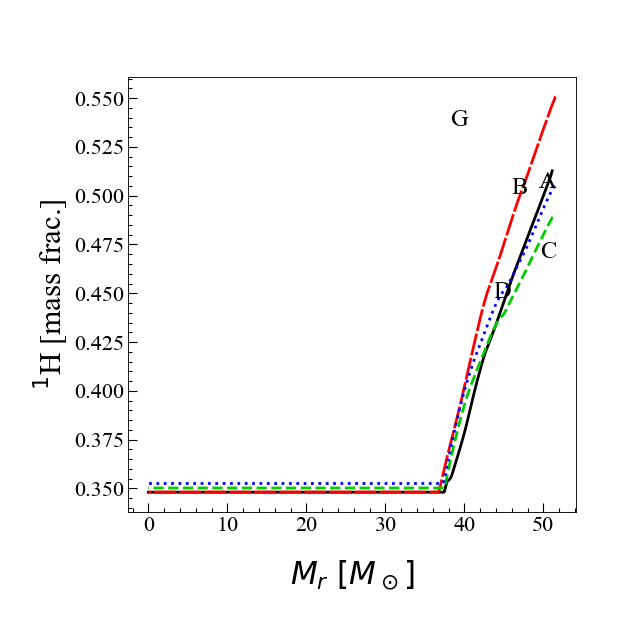}
        \includegraphics[width=0.33\textwidth]{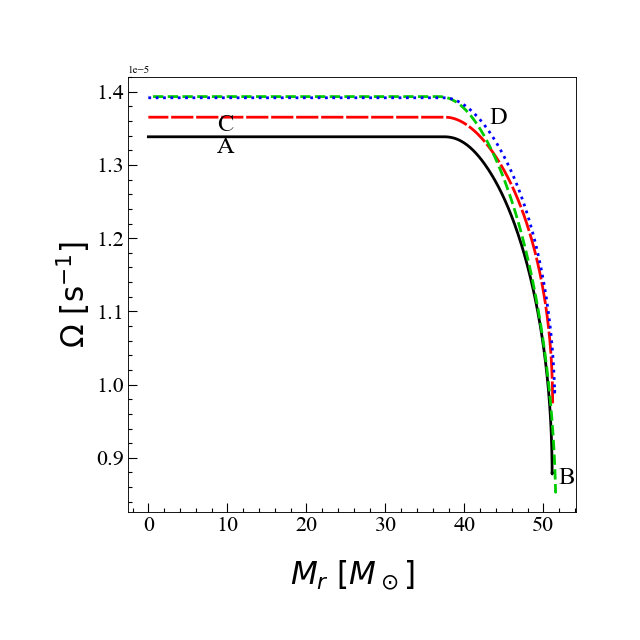}
      \caption{({\it Left panel}) Evolutionary tracks in the Hertzsprung-Russell diagram,({\it centre panel}) abundance of Hydrogen ranging from the center to the outer envelope of the models at Xc=0.35 versus the lagrangian  mass coordinate and ({\it right panel})
      the  variation of angular velocity   as a function of the Lagrangian mass coordinate in 60 M$_\odot$ at $Z$=0.014 and with 
$V_{\rm ini}/V_{\rm crit}=0.4$. The letters `A',`B',`C' and `D' and `G' correspond to the models as described in Table 1.}
         \label{fig_Z14_60_MS}
   \end{figure*}      

  \begin{figure*}
   \centering
    \includegraphics[width=0.24\textwidth]{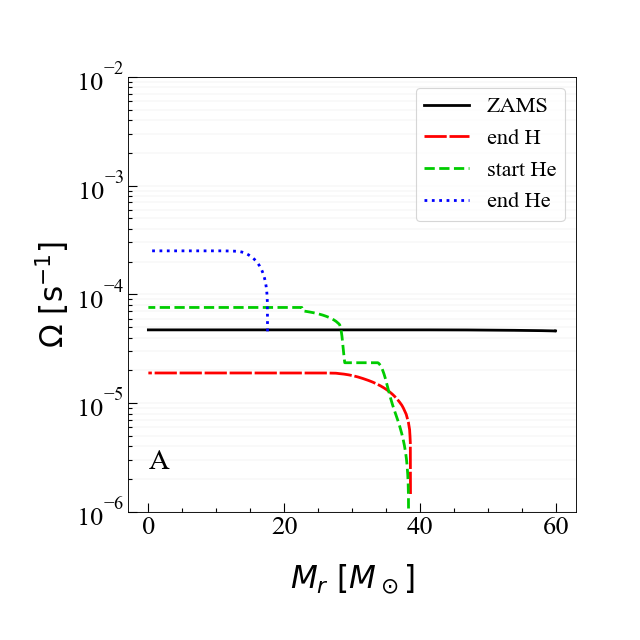}
    \includegraphics[width=0.24\textwidth]{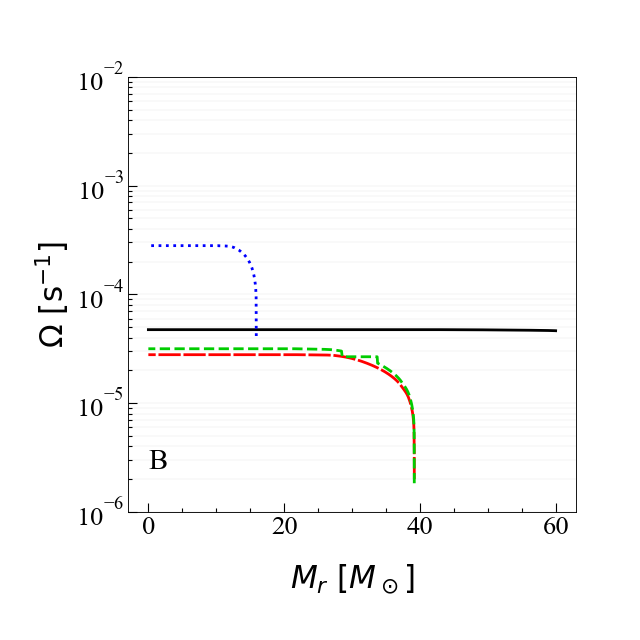}
    \includegraphics[width=0.24\textwidth]{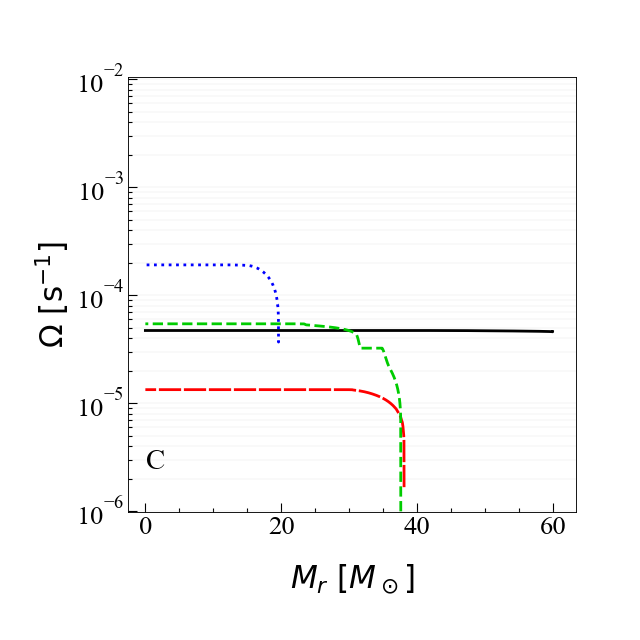}   
    \includegraphics[width=0.24\textwidth]{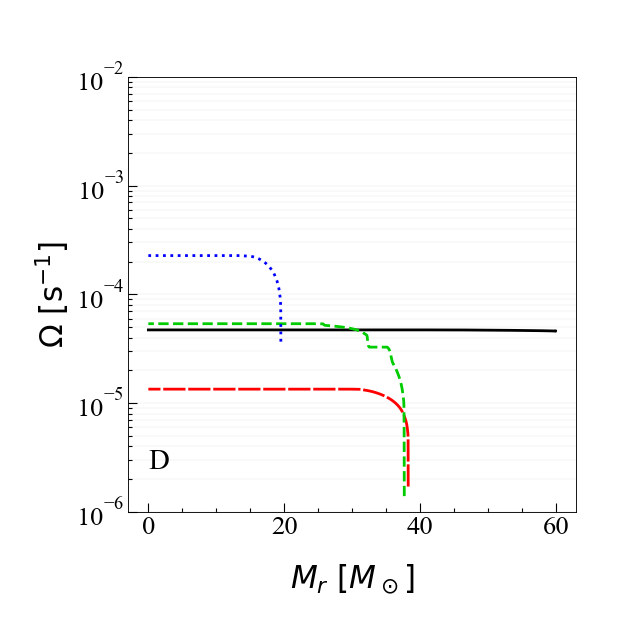} 

      \caption{Variation of angular velocity profiles as a function of mass for  60 M$_\odot$ for model A, B, C, D and C$*$ from left to right (see Table~\ref{tab:1}).}
         \label{fig_Omega_60ABCD}
   \end{figure*}   

The masses of the helium and carbon-oxygen cores at the end of the core He-burning phase are shown in columns 14 and 15 respectively of Table~\ref{tab:1}. The maximum relative difference in the helium core between rotating models, $(M_{\rm He}^{\rm max}-M_{\rm He}^{\rm min})/M_{\rm He}^{\rm min}$, is equal to 28\%. This is more than twice as large as the difference between a non-rotating model with a step overshoot of 0.25H$_{p}$ and without overshooting \citep[see e.g.][]{MM1987ov}.
Models with larger values of $D_{\rm h}$ show smaller helium cores at this stage. 
The maximum relative differences in the CO core and remnant masses between rotating models are $\approx 40$ and 17\% respectively\footnote{These relative changes are computed the same way as for the variations of the He-cores. For estimating the remnant mass, we have used a relation between the CO core mass and the remnant mass from \citet{Maeder1992}.}.
Overall, it is evident that the core sizes are significantly affected by choice of the prescription for rotation. The impact is greater than that of a moderate core overshoot in non-rotating models. These findings highlight the need for a better understanding of the underlying physics.

Figure~\ref{fig_Omega_jr_15_A} displays the variations of the angular velocity in different models during various evolutionary stages. The differences are minor at the end of the core H-burning phase, as the equations for computing the transport of angular momentum by the meridional currents are identical and are only weakly influenced by the choice of $D_{\rm h}$. However, changes to the chemical structure affect the tracks in the HR diagram, ultimately impacting the evolution of internal rotation. Indeed as seen in the left panel of Fig.~\ref{fig_Omega_H1He4c_15_14T}, the different models begins their core He-burning phase while the star is the blue or red part of the HR diagram. Because the stars is much more compact in the blue location than in the red one,
large differences in angular velocity are observed among the models at the beginning of the core-He burning phase.Additionally, we would like to remind the readers that the angular momentum transport and wind mass loss rates for massive stars are quite uncertain and degenerate with each other for many of the aspects discussed here. The actual masses at the end of the core He-burning phase vary by nearly 30\%. Regardless of the model considered, a huge ratio between the angular velocity of the core and that of the surface is observed at the end of the core He-burning phase, approximately 7-8 orders of magnitude. This indicates that fast rotating cores are a common feature in all models that only account for the hydrodynamical instabilities induced by rotation, {\it i.e.} that do not account for any magnetic instabilities.

\section{The results for the 60 M$_\odot$ models}\label{Sec:Results_60}   

\subsection{The diffusion coefficients}

At the middle of the core H-burning phase, the largest diffusion coefficient in all 60 M$_\odot$ models is  $D_{\rm eff}$ (Fig.~\ref{fig_D_60_14}).
This is in contrast to the 15 M$_\odot$ models, in which $D_{\rm shear}$ was the dominating diffusion coefficient throughout the star, except just above the convective core.
In the 60 M$_\odot$ models, a very flat rotation profile develops over a very large portion of the total mass of the star due the intrinsically large convective cores and the fact that mass loss removes the radiative differentially rotating layers (see the right panel of Fig.~\ref{fig_Z14_60_MS}).

Solid body rotation produces a large thermal imbalance because of large deformation in the outer layers that have a rotation near the one of the core, resulting in large values of the meridional current velocities.
To some extent these models exhibit a behaviour similar to the magnetic models (Sect.~\ref{Sec:Results_mag}) for the transport of the chemical species.
In the magnetic model, the flat profile results from the dynamo action.

\subsection{Tracks and lifetimes}\label{Sec:Tracks&Life}

The tracks for the 60 M$_\odot$ model and the distribution of hydrogen inside the model when the central mass fraction of hydrogen is 0.35,  can be seen in the left and middle panels of Fig.~\ref{fig_Z14_60_MS}. The profiles of hydrogen shown on the middle panel correspond to positions in the HR diagram at $\log L/L_\odot = 5.87$. Model B has the highest surface hydrogen abundance has the smallest effective temperature at this luminosity, while Model C has the smallest H surface abundance and the largest effective temperature. This is expected, since the closer a star approaches to chemically homogeneous structure, the bluer its position will be in the HR diagram. Note that the differences are small; less that 2\% and 5.5\% of differences in $\log L/L_\odot$ and in $\log T_{\rm eff}$ respectively.

The differences become more important toward the end of the MS phase. The convective core masses at a given mass fraction of hydrogen at the center are larger in models C and D than in models A and B when the mass fraction of hydrogen at the centre becomes smaller than 0.3. These larger cores allows these two models to reach higher luminosities at the end of the MS phase (see Fig.~\ref{fig_Z14_60_MS}).
The larger convective cores are due to larger $D_{\rm eff}$ in the models C and D. Interestingly, in these models, the values of $D_{\rm h}$ are actually smaller when the expression by \citet{Maeder2003} is used instead of that of \citet{Zahn1992}. 
This is an effect of the strong mass loss removing large amounts of angular momentum and thus decreasing $\Omega$.
We can see indeed that at the end of the MS phase, these models have lost more than 20 M$_\odot$.
The fact that $\Omega$ appears explicitly in the expressions of $D_{\rm h}$ by \citet{Maeder2003} and \citet{Mathis2004} implies that this quantity decreases as the evolution proceeds in the 60 M$_\odot$. This allows $D_{\rm eff}$ to increase.
This explains why in model C and D, the convective core masses at the end of the MS phase are larger than in models A and B.

\subsection{Surface velocities and abundances}

The surface velocities follow a similar evolution during the MS phase, regardless of the choice of prescription (see Models A to D in the third column of Table \ref{tab:1}). This is not the same for the surface abundances. Models C and D shows higher surface helium mass fractions than models A and B by about 20\%. Since all the models end with nearly the same total mass at the end of the MS phase, this difference is mainly a result of more efficient internal mixing inside Models C and D due to larger $D_{\rm eff}$. 

\begin{figure*}
   \centering
    \includegraphics[width=0.33\textwidth]{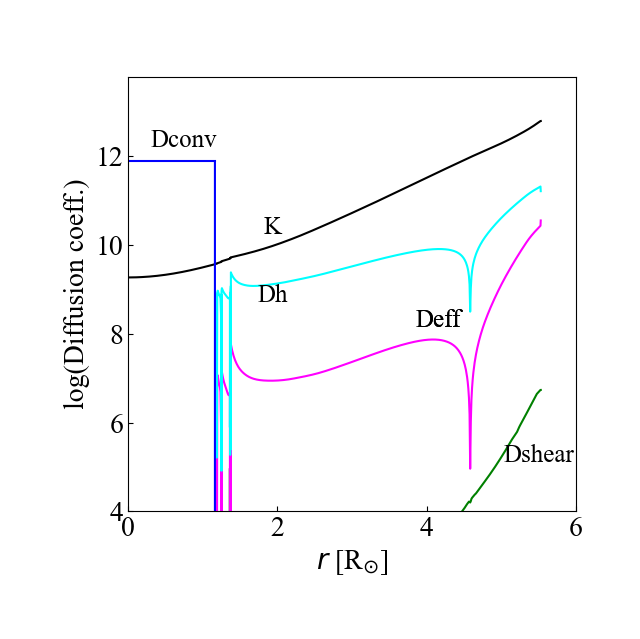} 
 \includegraphics[scale=.30, angle=0]{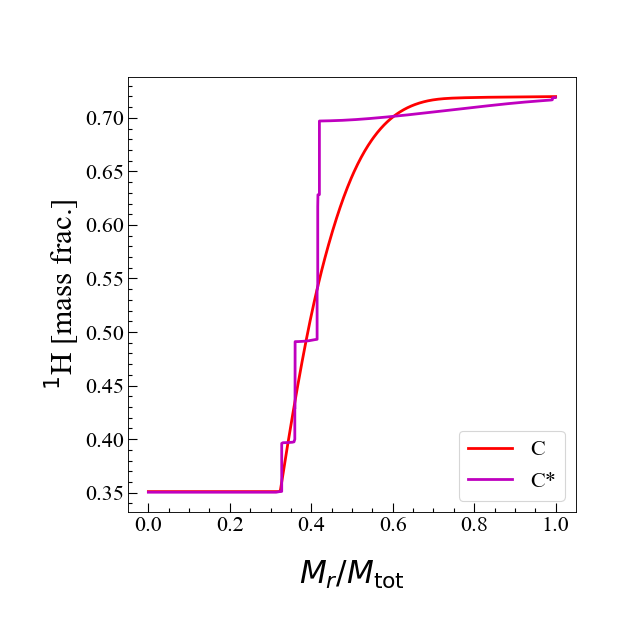} 
\includegraphics[width=0.33\textwidth]{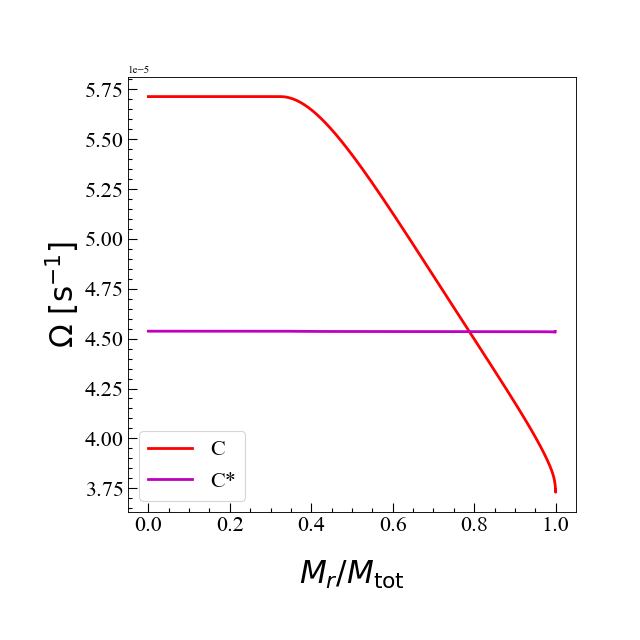}        
\caption{
{\it Left panel} Internal profiles of K the thermal diffusivity, 
      D$_{\rm conv}$ the convective diffusion coefficient, D$_{\rm shear}$ the shear diffusion coefficient, D$_{\rm h}$ the horizontal turbulence coefficient,  and D$_{\rm eff}$ the effective diffusivity in the 15 M$_\odot$ C$^*$ model with internal magnetic fields at solar metallicity. The quantities are plotted when the central mass fraction of hydrogen X$_{\rm c}$ = 0.35.
      {\it Middle panel} Variation of the H-mass fraction versus the lagrangian mass coordinate.
     {\it Right panel} Variation of the angular velocity versus mass in 15 M$_\odot$ for the C$^*$ and C models at the end of H-burning.
}
         \label{Fig_Dmago15}
   \end{figure*}

\begin{figure*}
   \centering
    \includegraphics[width=0.33\textwidth]{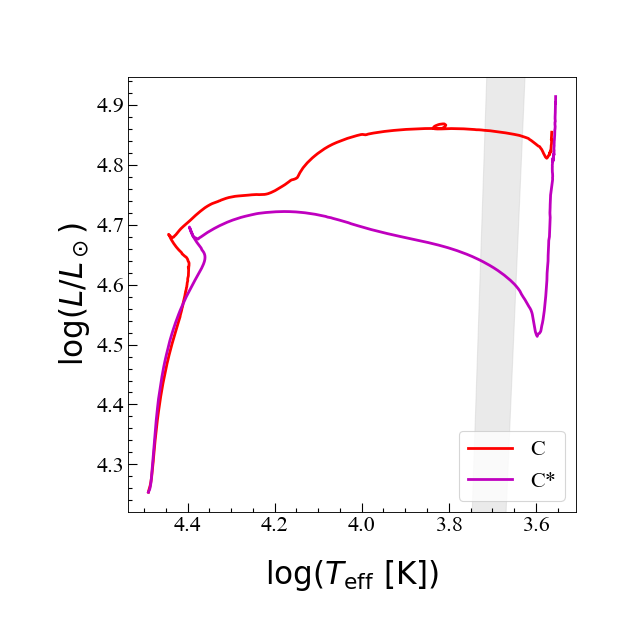} 
 \includegraphics[scale=.30, angle=0]{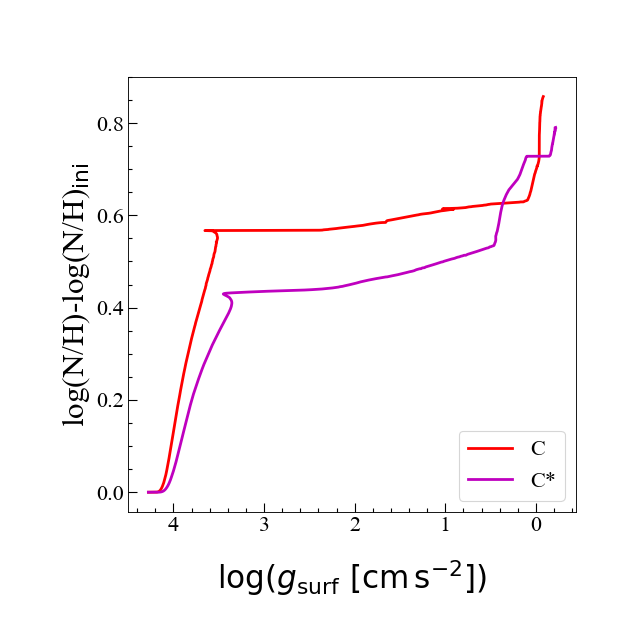} 
\includegraphics[width=0.33\textwidth]{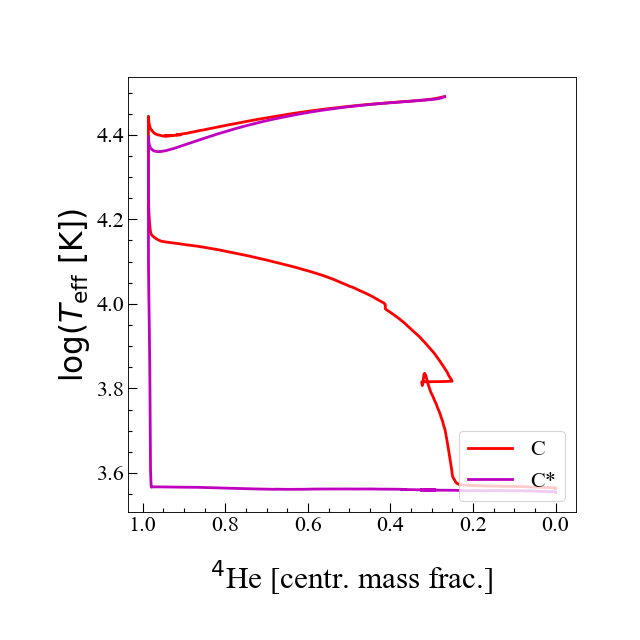}        
\caption{For models C and C$^*$ of 15 M$_\odot$ with 
$V_{\rm ini}/V_{\rm crit}=0.4$ are shown
{\it Left panel:}
The evolutionary tracks in the Hertzsprung-Russell diagram during the MS phase; 
{\it Middle panel:}
The change in the nitrogen over hydrogen ratio normalised to the initial value versus the surface gravity 
{\it Right panel:}
The variation of the effective temperature as a function of the central helium mass fraction.}
         \label{fig_M15_MS2}
   \end{figure*}   

\begin{figure}
\includegraphics[width=0.48\textwidth]{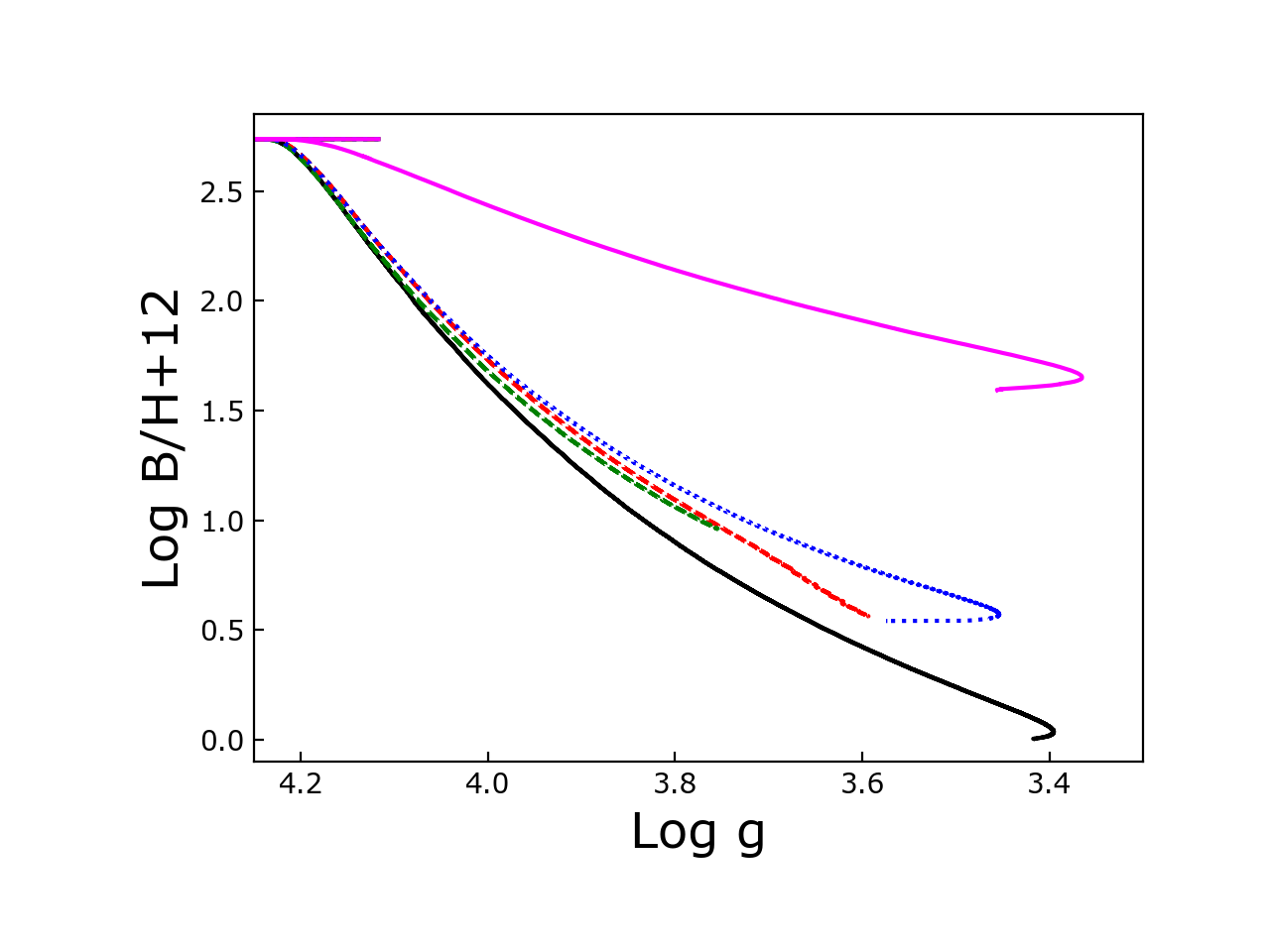} 
\caption{Evolution of the the B/H ratio (in number) at the surface of models A (continuous black curve), B (dashed green curve), C (dashed red curve), D (dotted blue curve) and C$^*$ (magenta continuous curve) models as a function of the surface gravity.}
\label{fig_B}
\end{figure} 

\begin{figure}
   \centering
    \includegraphics[width=0.24\textwidth]{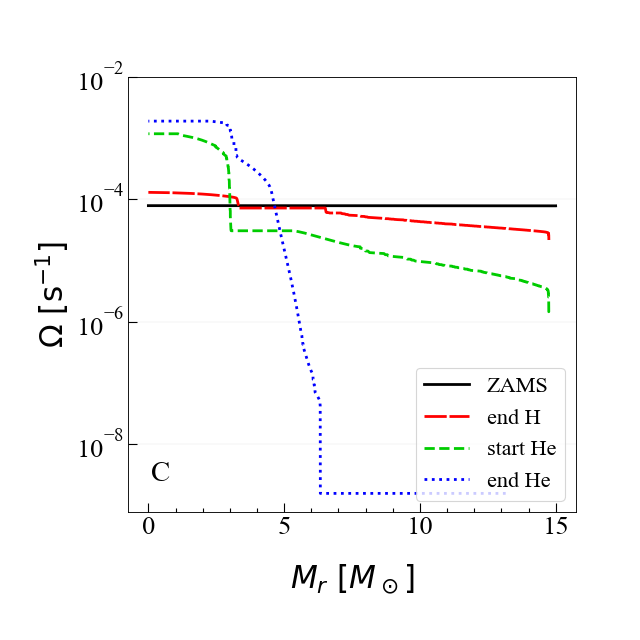}    
    \includegraphics[width=0.24\textwidth]{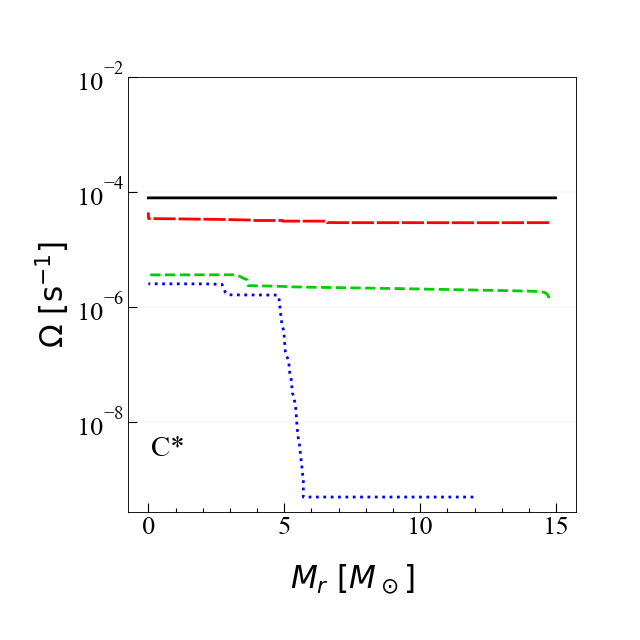}
      \caption{Variation of the angular velocity as a function of the lagrangian mass coordinate in 15M$_\odot$ at different evolutionary stages. The {\it left panel} is for the non-magnetic model C. The {\it right panel} shows the magnetic model C$^*$.}
         \label{fig_Omega_jr_15_magn}
   \end{figure}

\begin{figure*}
   \centering
        \includegraphics[width=0.33\textwidth]{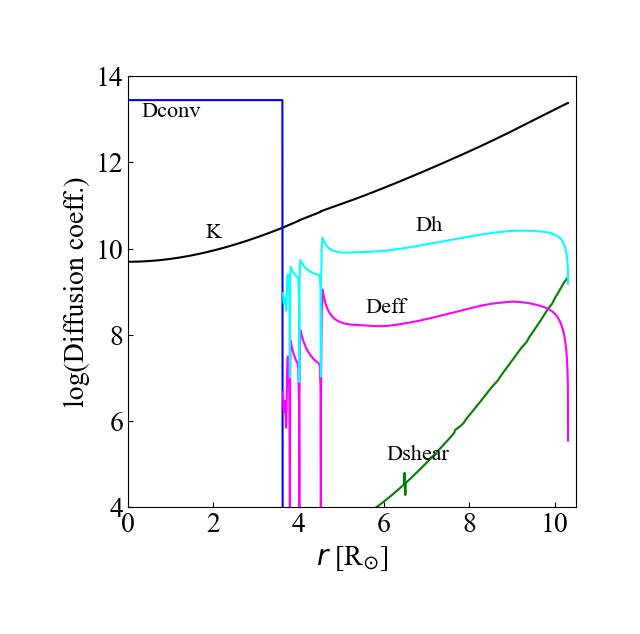}
    \includegraphics[width=0.33\textwidth]{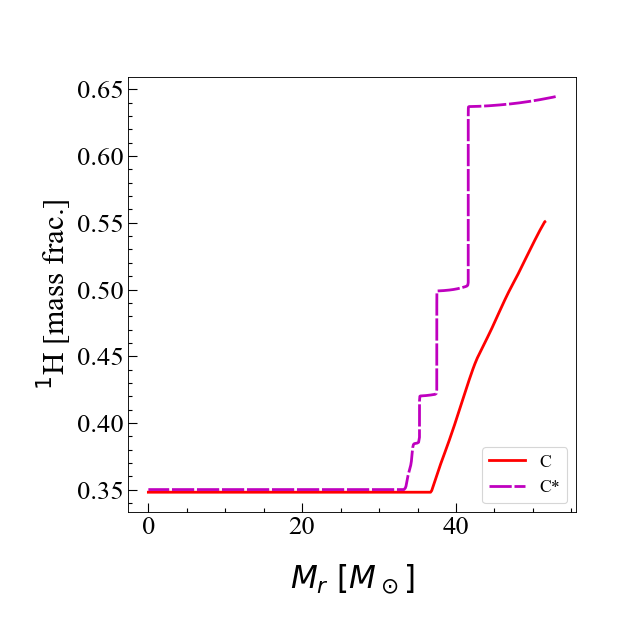}        
\includegraphics[width=0.33\textwidth]{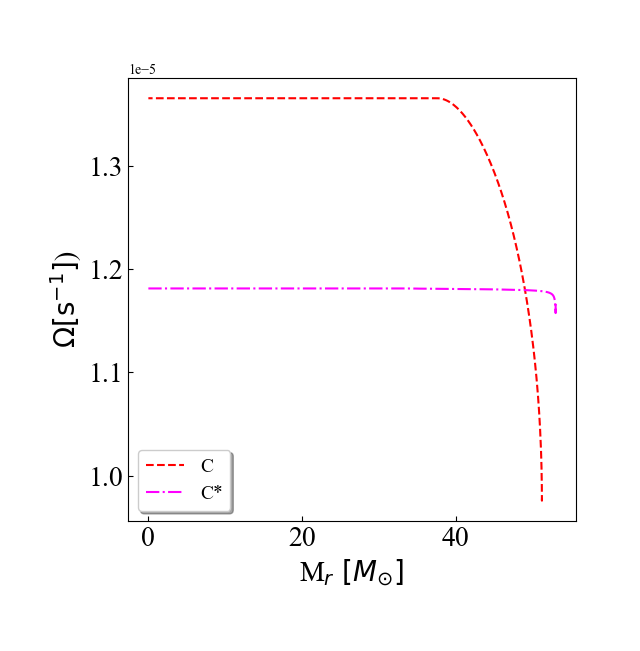}
      \caption{Same as Fig.~\ref{Fig_Dmago15} for 60 M$_\odot$ models.
}
         \label{Dmago60}
   \end{figure*} 

\begin{figure*}
   \centering
    \includegraphics[width=0.33\textwidth]{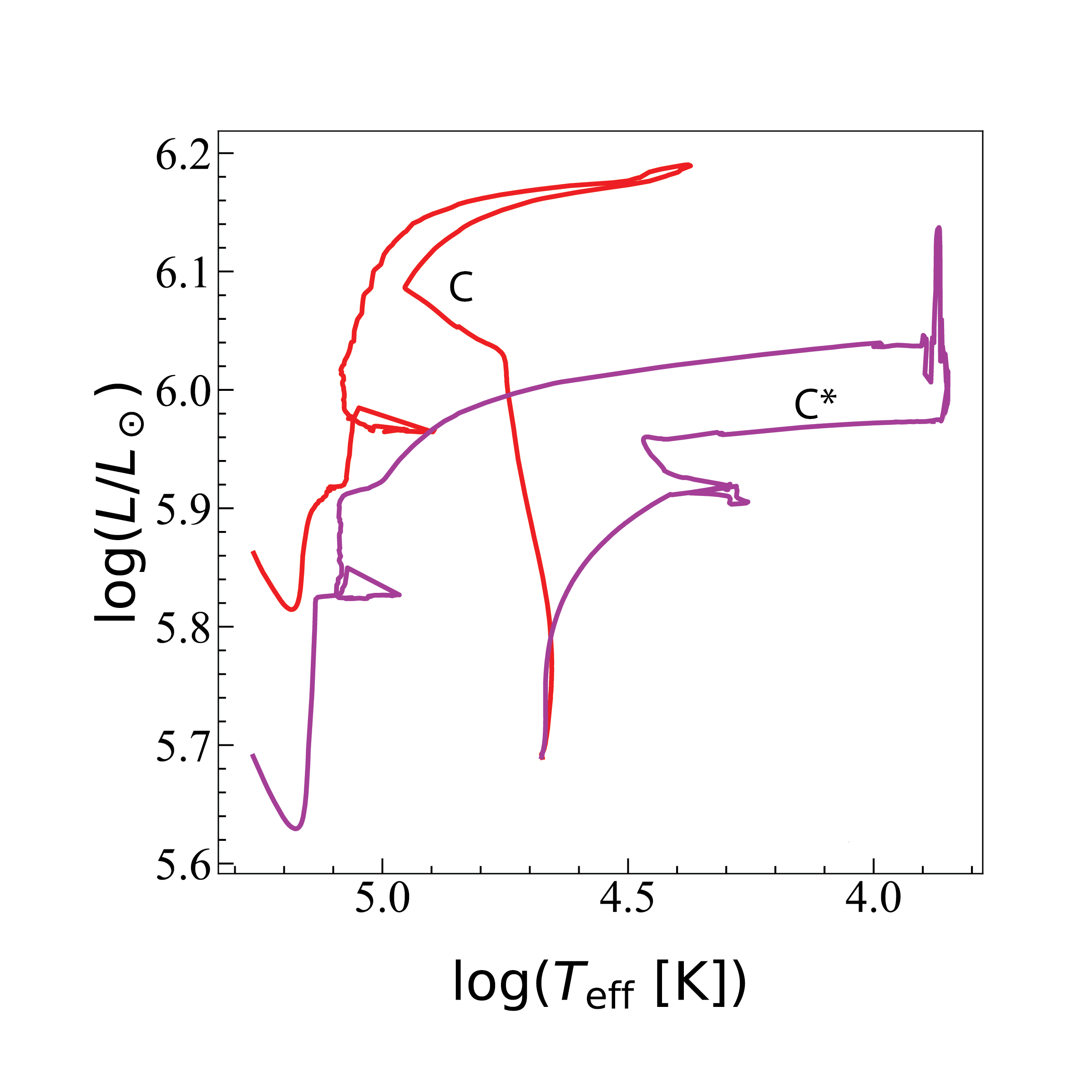} 
    \includegraphics[width=0.33\textwidth]{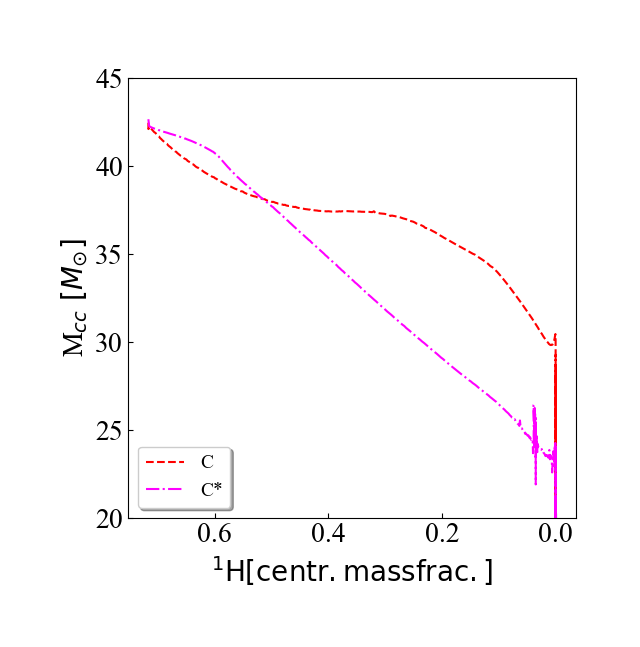}    
   \includegraphics[width=0.33\textwidth]{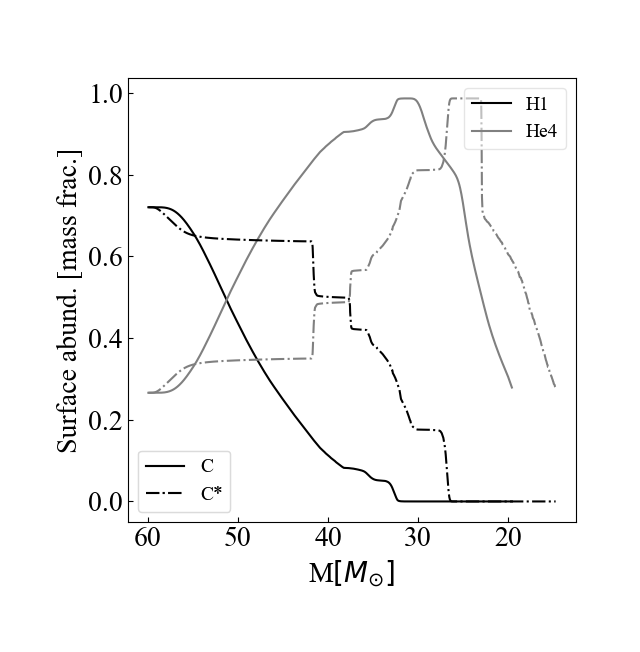}
    
\caption{Evolutionary tracks in the Hertzsprung-Russell diagram (left panel), variation of convective core mass versus the central mass fraction of hydrogen (center panel) and the surface abundances of hydrogen, helium, carbon, nitrogen and oxygen against the mass for models `C' and `C$^*$' as described in Table 1. 
}
\label{fig_M60_MSMT}
   \end{figure*} 

 \begin{figure}
   \centering
    \includegraphics[width=0.24\textwidth]{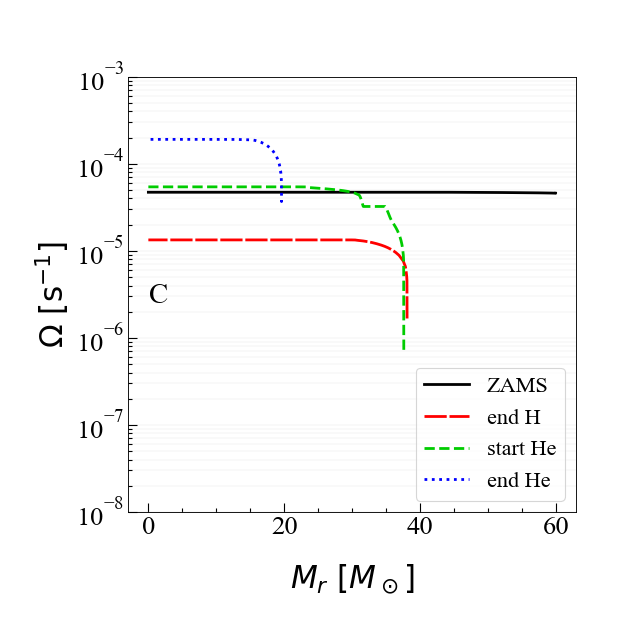}    
    \includegraphics[width=0.24\textwidth]{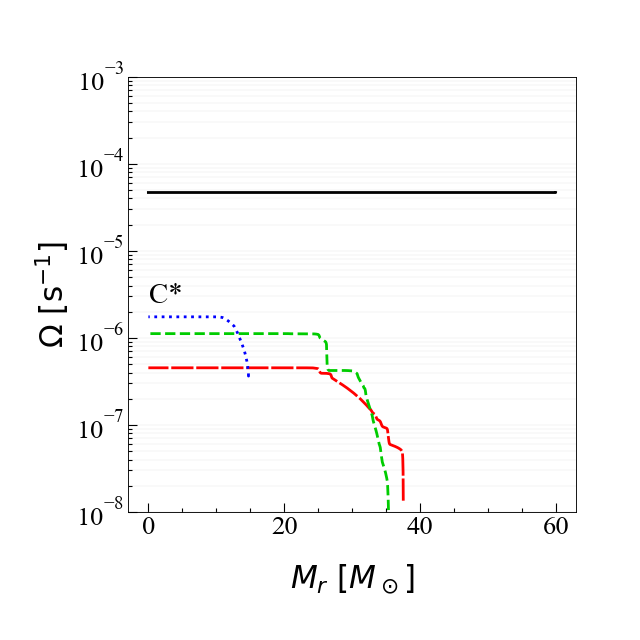}

      \caption{Same as Fig.~\ref{fig_Omega_jr_15_magn} for 60 M$_\odot$ models}
         \label{fig_Omega_60magn}
   \end{figure}

\subsection{Post MS evolution}

The non-rotating 60 \( M_{\odot} \) model evolves into a Wolf-Rayet star following the pathway described in the Conti scenario \citep{Conti1979}. In the case of the rotating model, the transition into the WR phase occurs earlier. This is due to the more efficient attainment of the necessary surface abundance changes for a WR star classification, facilitated by the mixing effects of rotation. It is important to note that these results are influenced by the underlying mass loss physics \citep[refer to][]{Fullerton2005, Gormaz2023}.

From the upper left panel of Fig.~\ref{fig_M60_MSMT}, we can see that the minimum luminosity reached during the WR phase for the non-magnetic models is between 5.65 and 5.8, corresponding to variations of  $<50$\% depending on the prescription used.

Models with an initial rotation of 0.4 $V/V_{\rm crit}$ have a shorter core He-burning lifetime by about 8 to 13\% compared to non-rotating models. The difference in lifetime between the different prescription models (5\%) is similar to the smallest difference between the rotating and the non-rotating models (8\%).
The core masses at the end of the core He-burning phase are indicated in Table 1. The more massive He-cores are those in models C and D (those with the stronger mixing). They differ from models A and B by at most 4 M$_\odot$ ($\sim 20$\%).
They also have larger CO core and remnant masses (assuming the star entirely collapses to a black hole). The differences are of 2-3 M$_\odot$ for the CO cores and slightly more than 1 M$_\odot$ for the remnant mass (also 20\% differences). While these differences are significant, these models will likely produce black holes, irrespective of the choice of rotational prescription \citep[see Fig. 3 in][]{Ertl2020}. In the context of this hotly debated topic, recent advancements in 3D neutrino-radiation hydrodynamics simulations have been significant, with multiple groups reporting on black hole (BH) formation accompanied by successful explosions and notable mass changes. Key contributions in this area include \citet{Ott2018}, \citet{Kuroda2018}, \citet{Chan2020}, \citet{Burrows2023} among others. For instance, \citet{Burrows2023} observed the formation of a 3 \( M_{\odot} \) BH from a 40 \( M_{\odot} \) progenitor. While these studies do not provide a definitive answer, they are essential for understanding the uncertainties related to rotational mixing in the context of explosion and collapse mechanisms. Currently, these uncertainties exceed 20\%, underscoring the need for comprehensive analysis in this field.

Figure~\ref{fig_Omega_60ABCD} shows the angular velocity inside the four models A, B, C and D at different stages of their evolution. The behaviors in all four models are very similar illustrating the fact that the changes of the diffusion coefficients have only an indirect impact on the internal angular velocity. In the 60 M$_\odot$ models, much of the evolution is driven by the stellar winds. The surface velocities during the Wolf-Rayet phase is small (of the order of 20-30 km s$^{-1}$) as a result of the large mass loss. Interestingly these models would produce very low spin black holes with $a_*\sim 0.003$ \footnote{The spin parameter is equal to $c J/(G M^2)$, where c is the velocity of light, $J$ the total angular momentum, $G$ the gravitational constant and $M$ the actual mass of the star at the end of the evolution. Here we consider values for the end of the core He-burning phase. Some mass and angular momentum can still be lost during the advanced phases and thus the value quoted for $a_*$ can be slightly different if a model at the presupernova stage was considered.}.

\section{Magnetic models}\label{Sec:Results_mag}

\subsection{The 15 M$_\odot$ model}

The diffusion coefficients for the 15 M$_\odot$ when the calibrated TS dynamo is used (Model C$^*$) are shown in the left panel of Fig.~\ref{Fig_Dmago15}. 
The expressions for $D_{\rm shear}$ and $D_{\rm h}$ in the magnetic model are identical to those in the non-magnetic model C. Therefore, in this section we will compare Model C$^*$ with Model C.
In Model C$^*$, the transport of angular momentum is dominated by the magnetic viscosity, while the transport of chemical elements is governed by the $D_{\rm eff}$ coefficient,
The shear diffusion coefficient has minimal impact due to its small magnitude. Consequently, in these models, changing the expression of $D_{\rm shear}$ has no effect on chemical element transport.

The middle panel of Fig.~\ref{Fig_Dmago15} illustrates the hydrogen profile at the middle of the core H-burning phase. When compared to the non-magnetic model (model C in red), the gradients above the core in the magnetic model are steeper and interspersed with nearly flat zones. This structure results from the receding convective core leaving around it a region of chemical gradients where some mixing can occur through the action of D$_{\rm eff}$. In those regions just above the core the mixing is not as efficient than in the region above where the meridional currents are stronger. We have thus an intermediate zone with a mixing timescale longer than the mixing timescale in the convective core and than in the radiative envelope. This tends to produce some staircase configurations. We would like to emphasize that the choice of the expression for $D_{\rm h}$ plays a key role. In the present paper, we present the case when the expression by \citet{Maeder2003} has been used. The expression of \citet{Zahn1992} would produce a much stronger chemical mixing. These aspects will be discussed in forthcoming papers.

The left panel of the same figure shows that $D_{\rm eff}$ is much larger in the Model C$^*$ than $D_{\rm shear}$ in model C (Fig.~\ref{fig_15_DIFF}). As a result, in the magnetic model, hydrogen tends to move inward and accumulate above the convective core due to the decrease of $D_{\rm eff}$ just above the core, leading to steeper gradients.
The high viscosity value results in a flat rotation profile during the middle of the MS phase, as evidenced by the right panel of Fig.~\ref{Fig_Dmago15}.

During the MS phase, the magnetic model of the 15 M$_\odot$ star has a wider range of effective temperatures than its non-magnetic counterpart, owing to a lower degree of mixing (left panel of Fig.~\ref{fig_M15_MS2}). This weaker mixing is apparent in the evolution of surface abundances, as shown in the middle panel of Fig.~\ref{fig_M15_MS2}.
The evolution of the effective temperature as a function of the mass fraction of helium in the centre is shown in the right panel of Fig.~\ref{fig_M15_MS2}. The magnetic model rapidly becomes a red supergiant just after the MS phase, while the non-magnetic model burns a large part of its helium at high effective temperatures. As discussed earlier, small changes in the abundances profiles around the H-burning shell are responsible for this difference \citep{Walm2015, Klencki2022, Eoin2022}. This sensitivity on small changes in the He-distribution may question the robustness of this result to changes in the space and time discretization. It would go well beyond the present work to discuss in detail this point and the reader must therefore keep in mind this possibility.

In Fig.~\ref{fig_B}, we show the variation of the boron abundance at the surface of the A, B, C, D and C$^*$ models. As briefly mentioned in the introduction, boron does not need to be dredged down very deeply inside the star to be affected by nuclear reactions.
It is is destroyed via proton capture at temperatures of around 6 $\times$ 10$^6$ K \citep[see e.g.][]{Proffitt1999}, making it a good indicator of mixing in the outer layers of stars, where $D_{\rm shear}$ dominates in non-magnetic models and $D_{\rm eff}$ in magnetic ones. Despite some differences, models A, B, C, and D exhibit similar qualitative evolution, largely due to the fact that the depletion of boron occurs primarily in zone 4, where the two $D_{\rm shear}$ expressions yield comparable values. Nonetheless, the depletion of boron in magnetic models is weaker than in non-magnetic models (at a fixed initial mass, rotation and metallicity), which could be used to differentiate between the two types of models. Further research is necessary to investigate this possibility, but it is clear that magnetic models exhibit distinct behaviors in the boron versus log g, boron versus nitrogen surface abundances, and boron versus surface rotation planes.

In Figure~\ref{fig_Omega_jr_15_magn}, we compare the angular velocity distribution of a non-magnetic model (left panel) and a magnetic model (right panel). At the end of the core He-burning phase, the magnetic model displays a core angular velocity three orders of magnitude lower than the non-magnetic model. This difference suggests that the angular momentum of the remnant would decrease by three orders of magnitude, resulting in a proportional increase in its spin period; assuming the angular momentum of the part of the star that becomes the neutron star remains constant post-He-burning.

\subsection{The 60 M$_\odot$ model}

The left panel of Fig.\ref{Dmago60} illustrates the diffusion coefficients in the magnetic 60 M$_\odot$ model, revealing that the same trends observed in the corresponding 15 M$_\odot$ model also hold qualitatively in the 60 M$_\odot$ model. Throughout the radiative zone, angular momentum transport is dominated by the magnetic viscosity, while chemical species transport is dominated by $D_{\rm eff}$. There is only a small region near the surface where $D_{\rm shear}$ exceeds $D_{\rm eff}$ due to the differential rotation present in the outermost layers (as depicted in the right panel of Fig.\ref{Dmago60}).

The profile of the hydrogen abundance at the middle of the core H-burning phase is shown in the middle panel of Fig.~\ref{Dmago60}. The magnetic model retains an H-rich outer envelope with a steep gradient connecting it to the convective core.
This behavior mirrors that of the magnetic 15 M$_\odot$ model (as depicted in the middle panel of Fig.~\ref{Fig_Dmago15}). The right panel of Fig.~\ref{Dmago60} shows the angular velocity variation with respect to the Lagrangian mass coordinate at the middle of the core H-burning phase, with the magnetic model exhibiting a core that rotates more slowly (by roughly 15\% compared to the corresponding non-magnetic model) and displaying less differential rotation in the outermost layers.

The impact of magnetic instabilities on the 60 M$_\odot$ evolutionary track is evident in the left panel of Fig.~\ref{fig_M60_MSMT}. Model C, which undergoes strong internal mixing, evolves almost vertically and remains in the blue, while the magnetic model evolves towards the red. As depicted in the middle panel of Fig.~\ref{fig_M60_MSMT}, the convective core is significantly larger in the non magnetic model compared to the magnetic one for most of the core H-burning phase.
The right panel of Fig.~\ref{fig_M60_MSMT} presents the surface abundance evolution as the actual mass of the 60 M$_{\odot}$ star decreases. The hydrogen mass fraction at the surface becomes smaller than 0.4 when the actual mass of the star is around 35 M$_\odot$ in the magnetic model and around 50 M$_\odot$ in the non-magnetic model. 
Figure~\ref{fig_Omega_60magn} compares the angular velocity variations at different evolutionary stages between the 60 M$_\odot$ non-magnetic (left panel) and magnetic (right panel) models. The strong coupling imposed by the magnetic field significantly slows down the core's rotation in the magnetic model compared to the non-magnetic model, reducing it by two orders of magnitude at the end of the core He-burning phase.

\section{Discussion and conclusions}\label{Sec:Disc_Conc}

In this study, we have examined the impact of changing the expressions for the diffusion coefficients for rotation in our stellar models, while keeping all other physical inputs unchanged. Our findings demonstrate that the specific choice of the expressions for $D_{\rm shear}$ and $D_{\rm h}$ in non-magnetic models can yield markedly distinct outcomes for a given initial mass, rotational velocity, and composition. Furthermore, there are significant variations between the magnetic and non-magnetic models. We enumerate our key findings below.

\begin{enumerate}
\item Altering the diffusion coefficients in non-magnetic models does not have a significant effect on angular momentum transport, but rather impacts the star's evolution by altering its chemical structure.
\item Regardless of the chosen diffusive coefficients or initial mass, non-magnetic models tend to produce very rapidly rotating cores at the end of the core He-burning phase, consistent with results by \citet{Georgy2009}.
\item In magnetic models, the magnetic instability dominates angular momentum transport and is largely unaffected by the choice of $D_{\rm shear}$ and $D_{\rm eff}$. The models exhibit flat rotation profiles during the main sequence phase for 15 and 60 M$_\odot$ stars. Magnetic models result in significantly slower rotating cores at the end of the core He-burning phase, consistent with previous findings by \citet{Heger2005} and \citet{FullerMa2019}, although their magnetic transport of angular momentum is implemented differently.
\item The choice of diffusion coefficients has a significant impact on the mixing of chemical elements and the evolutionary tracks in the HR diagram. In non-magnetic models, changing the expression for $D_{\rm shear}$ affects the evolution of the 15 M$_\odot$ model but not the 60 M$_\odot$ model. Using the expression for $D_{\rm shear}$ from \citet{Maeder1997} instead of \citet{TZ1997} produces more luminous and bluer evolutionary tracks for the 15 M$_\odot$ model, as well as stronger surface enrichment at the end of the MS phase.
\item Increasing $D_{\rm h}$ in the non-magnetic 15 M$_\odot$ model reduces the transport just above the convective core, where $D_{\rm eff}$ dominates. The expression for $D_{\rm h}$ from \citet{Maeder2003} instead of \citet{Zahn1992} produces smaller convective cores, lower luminosity tracks, and less mixed models at the end of the MS phase.
\item Reducing $D_{\rm eff}$ generally leads to a decrease in mixing, but this decrease may not be significant if high values of $D_{\rm shear}$ are present in the radiative envelope above the region directly surrounding the core with strong chemical gradients during the core H-burning phase. This is due to the mild chemical gradients resulting from the retreating convective core during the early phase of the core-H-burning phase.
\item The results of the current computations are consistent with previous {bf studies by \citet{MMVII2001, Walm2015, Klencki2022} and \citet{Eoin2022}, which found} that the beginning of the core He-burning phase occurs as a red supergiant if the region near the H-burning shell is enriched in helium. The timescale for the star to cross the HR gap after the MS phase plays an important role in determining the blue to red supergiant ratio and has implications for close binary evolution, particularly for case B mass transfer.
\item The differences in MS lifetimes among the rotating models computed with different diffusion coefficients are as significant as the differences between rotating and non-rotating models computed with and without a moderate core overshoot.
\item The masses of the cores at the end of He-burning also presents significant differences depending on the prescriptions used. These differences may amount to twice the differences when comparing against non-rotating models with and without a moderate overshoot.
\item The 60 M$_\odot$ model is strongly affected by mass losses through stellar winds. Changing the diffusive coefficients has only a weak impact.
\item The primary process responsible for the mixing of chemical elements in non-magnetic 60 M$_\odot$ models is $D_{\rm eff}$. 
Convection and mass loss cause the star to quickly transform into a nearly solid body rotating core, leading to increased transport of chemical elements through meridional currents and consequently through $D_{\rm eff}$.
\item In models using expressions of $D_{\rm h}$ that show an explicit dependence on $\Omega$ and undergo significant mass loss, the value of $D_{\rm eff}$ increases with time, leading to the development of larger convective cores at the end of the MS phase. This is because in these models, $\Omega$ decreases with time due to mass loss and as a result, $D_{\rm h}$ decreases and $D_{\rm eff}$ increases.
\item Chemical mixing in magnetic models is controlled by $D_{\rm eff}$, which is affected by the choice of $D_{\rm h}$ but not by $D_{\rm shear}$ when the rotation profile is very flat. For this study, only magnetic models using the $D_{\rm h}$ from \citet{Maeder2003} are presented. The degree of mixing in both the 15 and 60 M$_\odot$ magnetic models is lower than that of the corresponding non-magnetic models using the same $D_{\rm h}$.
 
\end{enumerate}

The aforementioned changes have an impact on the reproduction of the observed width of the MS band by a step overshoot. For instance, a smaller value of step overshoot is required when using the $D_{\rm shear}$ from \citet{TZ1997} when compared to \citet{Maeder1997}. However, the calibration of convective boundary mixing must be done in conjunction with the calibration based on the chemical surface enrichment during the MS phase, as the size of the convective cores also affects these changes. Model A employs the same physics as the more recent grids of stellar models by the Geneva group \citep{Eks2012, Georgy2013, Groh2019, Murphy2020, Eggenberger2021, Yusof2022}. The surface abundances shown by Model A during the MS phase provide a reasonable match to the observed surface enrichment of stars with an average rotation velocity \citep[as discussed in][]{Eks2012}. 
The right panel of Fig.~\ref{fig_Z14_CCNH} indicates that the models C and E would require minimal adjustments if they use $D_{\rm shear}$ from \citet{Maeder1997} (see middle panel of Fig.~\ref{fig_Z14_CDEF_222}).
However, models B, D, and F, which use $D_{\rm shear}$ from \citet{TZ1997}, would require either higher initial rotations or an increase in their diffusion coefficients by selecting a value for $f_{\rm en}$ greater than 1.0 to achieve a similar enrichment as model A.

Although determining the optimal physics from surface observations alone remains a challenge, asteroseismic analyses of massive and low-mass stars provide valuable insights into angular momentum transport \citep[e.g.,][]{Beck2012, Mosser2012, Deheu2012, Gehan2018, Deheu2020, BETA2022, Pedersen2022}, while helioseismology is useful for understanding the Sun's internal rotation \citep{Couvidat2003}. Currently, magnetic models seem to be more effective in reproducing the internal rotation of the Sun \citep{Sun2019, NatSUN2022}, as well as that of low-mass subgiants, red giants, and clump stars \citep{TESZS2019, FACUNDO2022}. For massive stars, fewer constraints of this nature are available, although asteroseismic data on slowly rotating $\beta$-Cephei stars has produced varied results \citep{BETA22022, BETA2022}. Some stars show internal angular velocity gradients that cannot be explained by existing magnetic models, while others exhibit a flat rotation profile internally. The spin of compact remnant favors very efficient angular momentum transport \citep{Heger2004, WD2019}.
If both solid-body and differential rotation are observed in nature, the key challenge lies in elucidating the physical mechanisms that give rise to these two distinct rotational behaviors.

\acknowledgement{The authors would like to thank the referee for the insightful comments and contributions towards enriching this paper. We would also like to thank Yves Sibony for sharing his results on the impacts of changing the time resolution. DN, GM, SE, FDM, PE, CG, EF have received funding from the European Research Council (ERC) under the European Union's Horizon 2020 research and innovation programme (grant agreement No 833925, project STAREX). This work was supported by the Fonds de la Recherche Scientifique-FNRS under Grant No IISN 4.4502.19.}

\bibliographystyle{aa}
\bibliography{Z014}

\appendix

\section{A brief discussion of the physics of rotation}\label{Sec:ApA}

For the sake of completeness, we remind here a few points about the physics of rotation included in the present models. In models without magnetic fields, the angular momentum is mainly transported by meridional currents, while the chemical species are transported by shear instabilities and meridional currents.

The transport of the angular momentum by the meridional currents is computed resolving an advecto-diffusive equation (see Eq.~2) during the MS phase\footnote{After the MS phase, a diffusive approach is applied because the structure becomes too complexe with intermediate convective zone for allowing to resolve the advecto-diffusive equation. This is however not a too serious shortcoming since the timescales during the post MS phases are reduced and one of the dominant effect governing the change of the angular velocity inside the star in the hydrodyanmical model is the local conservation of the angular momentum.}. In some models the transport of the angular momentum by the meridional currents is computed resolving a diffusive equation \citep[see e.g.][]{Heger2000,Chieffi2013, Choi2016, Nguyen2022}, in other the advecto-diffusive equation has been resolved \citep{Talon1997, MM2000, Palacios2003, Chieffi2013}. Note that \citet{Chieffi2013} have implemented both the advecto-diffusive and the diffusive equation for the transport of the angular momentum, but most of the results presented in this paper have been obtained with the diffusive approach. Let us remind here a few points: first the transport by meridional currents is an advective process and thus in principle it should be accounted for by resolving such an equation. It happens that
using a diffusive equation predicts a much flatter distribution of the angular velocity during the MS phase as can be seen looking at Fig.~1 in \citet{Griffiths2022}. 

The present models explicitely account for an expression for the strong horizontal shear diffusion $D_{\rm h}$
\citep{Zahn1992, Maeder2003, Mathis2004}. Some authors \citep[see e.g.][]{Heger2000, Choi2016, Nguyen2022} do not explicitely account for an explicit expression for $D_{\rm h}$ although they implicitely account for it assuming that a process is responsible for the shellular rotation state that any 1D stellar evolution model with rotation assumes. Interestingly $D_{\rm h}$ is involved in the expression of $D_{\rm shear}$ of \citet{TZ1997}
as a factor reducing the inhibiting effect of the chemical stratification. Thus it replaces the use of a constant value parameter (often called $f_\mu$) that multiplies the $\mu$-gradients, {\it i.e.} that replace $\nabla_\mu$ by $f_\mu \nabla_\mu$ with $f_\mu$ having a value of the order of the percent in the criterion that gives the lower limit of the shear enabling the instability to be activated \citep{Heger2000}. The value of this parameter is chosen in order for models with an averaged surface rotation compatible with the observations  to reproduce some surface nitrogen enrichment at the end of the MS phase \citet[see e.g. section 3 in][]{Chieffi2013}. Another effect of considering an explicit expression for $D_{\rm h}$ is that it naturally allows to account for the different efficiencies of the meridional currents in transporting the angular momentum and the chemical species. Indeed this comes from the expression of $D_{\rm eff}$ (see Eq.~8). Note that in this expression the factor 1/30 is not a free parameter it comes simply from the computation of $\int_0^\pi |P_2(\cos \theta)|^2 \sin \theta {\rm  d} \theta$, where $P_2(\cos \theta)=0.5(3 \cos^3 \theta -1)$ \citep{Chaboyer1992}.
Note that for the transport by the shear, the efficiency appear to be the same for the angular momentum and the chemical species according to the simulation by \citet{Prat2016}.

In the present approach, the free parameters are on one side the fraction of the energy used for driving the transport and the numerical parameters in front of the expressions of the strong horizontal diffusion. For the first, we chose to consider here that the whole energy is used for triggering the mixing. Actually this parameter can be interpreted in two ways, either as the fraction  of the excess energy in the shear that can be used for the transport for a given critical Richardson number
\footnote{The Richardson number is a dimensionless number defined as the ratio of the buoyancy term to the flow shear term. The critical value is the minimum value for the medium to become turbulent. The value of 0.25 is explained in for instance in \citet{Shu1992}} or as a value for the critical Richardson number given a fraction of the energy in the shear that can be used for the transport \citet[see Eq. 2.2 in][for the expression of the Richardson criterion]{Maeder1997}. Whatever the choice done here, there is no reason to change the assumption depending on the expressions used provided they are based on the same general physical assumptions which is the case here.We adopted the numerical parameters in front of the $D_h$ expressions as suggested by the authors that have proposed this expressions, since these authors chose these numerical factor based on physical arguments. Thus, we are left with a completely determined system allowing to compute models that can be compared because based on similar physical assumption.

For the magnetic models we have that the main mechanism for the transport of the angular momentum is the magnetic viscosity, as described in the works of \citet{Spruit1999, Spruit2002, HWS2005}. In those models we did not account for the angular momentum transport by the meridional currents. Actually meridional currents have little effect on the redistribution of the angular momentum. These currents would most of the time transport angular momentum from the outer regions of the star into the inner regions and accounting for them requires the use of very small time steps when applied to fast rotating massive stars \citep{MMB2005}. In the present models we chose to not account for this transport. In those models the mixing of the chemical elements is mainly driven by $D_{\rm eff}$ thus the concomitant effect of meridional circulation and the strong horizontal shear turbulence. The Tayler-Spruit calibrated Magnetic models have as free parameters the value $C_{\rm T}$ and (as for the non-magnetic models), the numerical parameters in front of the expressions of the strong horizontal diffusion. Here we present models with the expression and the numerical values given in \citet{Maeder2003}.

Finally let us add that all the models begins their evolution with a flat profile of $\Omega$ at the ZAMS. In case of non-magnetic models, the initial distribution of the angular velocity very rapidly converges towards an equilibrium profile so that the inwards transport by merdional currents is balanced by the outwards transport by shear \citep[see the discussion in][]{Denissenkov1999, MM2000}. It has also been discussed in \citet{Granada2014}, that starting on the ZAMS with a differential rotation profile or from a solid body one gives a similar evolution as long as the total angular momentum of the model is the same. In case of magnetic models, the coupling due to the magnetic instability is so strong (especially at the beginning when there is no inhibiting $\mu-$gradients) that starting from different profiles from a solid body one would not change much the fact that star will have after a short time a flat rotation profile.

\section{Physical ingredients of the stellar models}\label{Sec:ApB} 

As said above we adopted the same physical ingredients as in \citet{Eks2012}.
We just remind here some main properties. The models are computed with a step overshooting. The radius of the convective core is given by
$R_{\rm Schw}+0.1 H_p$, where $R_{\rm Schw}$ is the radius of the convective core given by the Schwarschild criterion for convection and $H_p$ is the pressure scale heigh~t estimated at that position. We account for such a step overshooting during the core H and He-burning phases. We did not account for any overshooting for intermediate convective zones or at the bottom of the outer convective zone. 

The mass loss rate scheme is described in details in Sect.~2.6 of \citet{Eks2012}. During the MS phase, the prescription by \citet{Vink2001}
is used. In the domain not covered by this prescription, we have used the fit proposed by \citet{deJager1988}. Note that this last prescription implicitely account for an increase of the mass flux when the model enters into the observed luminosity and effective temperature domain corresponding to Luminous Blue Variables. For red supergiants with $\log T_{\rm eff} < 3.7$, a linear fit between the data by \citet{Sylvester1998} and those of \citet{vanLoon1999} has been used. During the Wolf-Rayet phase, the prescription by \citet{Nugis2000} has been adopted. In the small domain where the prescription by \citet{Grafener2008} is valid, it was preferred to that of \citet{Nugis2000}. The correcting factor due to rotation is accounted for as proposed by \citet{MM2000}.

No thermohaline mixing or semiconvection is considered. Chemical elements 
are homogeneously mixed in convective zones. In rotating models, we have some mixing also occurring in radiative regions induced by rotation.

The time steps are controlled by the nuclear energy generation inside the star and additionally is limited by the change of major physical quantities such as central fuel abundance, T$_{eff}$, $\rho_{c}$, critical rotation and transport of $\Omega$. The space resolution is chosen so that the gradients with respect to the Lagrangian mass coordinate  in all the main structure quantities as the pressure, temperature, luminosity and radius are below some predefined values.

\section{Resolution tests}

\begin{figure}
   \centering
    \includegraphics[width=0.42\textwidth]{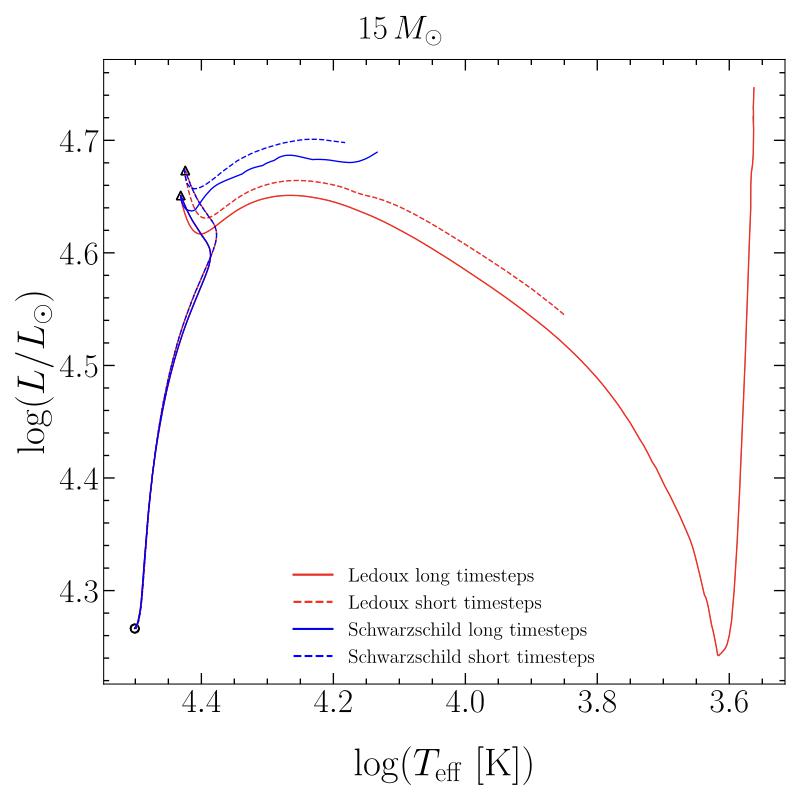}    

      \caption{HR diagram for four (Ledoux and Schwarzschild, long and short timescales) 15 M$_\odot$ models without rotation, showing a comparison of the effect of time resolution on evolution.}
         \label{fig_ApB}
   \end{figure}

This appendix presents our findings from the resolution tests conducted using HR diagrams. We studied four 15 \( M_{\odot} \) models without rotation, as shown in Figure~\ref{fig_ApB}. These models are crucial for understanding numerical convergence in computational stellar physics. There are two groups of models based on their timestep sizes. The first group, with the largest timesteps, is shown in solid lines. This group's evolution is traced until the onset of core helium-burning. The second group, with timesteps 10 to 100 times smaller, is shown in dashed lines. For this group, we only trace the evolution up to just past the main sequence due to long computational times and early divergence. Analysis of these models shows that changing the timestep size alters the blue turn-off's effective temperature and luminosity by about 0.02 dex. 
Therefore any differences smaller than this between
models computed with different physics is likely irrelevant beacuse it can be due to difference in discretization rather than caused by a difference in the physics.
These changes highlight the limitations of present computations.

\section{Smoothing of d$\Omega$/dr}

Meridional circulation in GENEC implies the fourth order derivative of $\Omega$ and this is obtained by performing a polynomial fit on a sliding window. The polynomial itself is of second order and we use the NAG subroutine e02acf to calculate it. The size of the window is determined by the number of shells max (5, number of shells/120). This fitting of the run of $\Omega$ is only used for computing the derivative of $\Omega$ with respect to $r$ but is of course not replacing the actual value of $\Omega$ obtained from the resolution of the transport equations.

\section{Figures for prescriptions E and F}

We discuss here the models computed with the $D_{\rm h}$ of \citet{Mathis2004} and compare them with the models computed with the $D_{\rm h}$ of \citet{Maeder2003}.

\begin{figure*}
   \centering
    \includegraphics[width=0.33\textwidth]{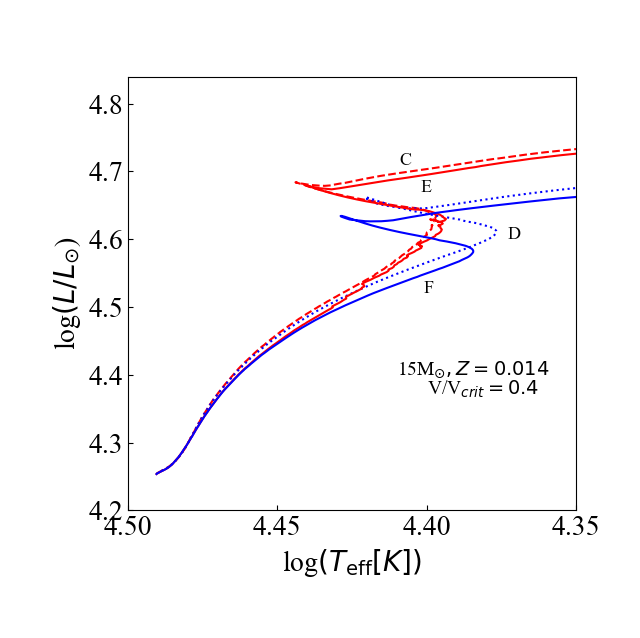}
    \includegraphics[width=0.33\textwidth]{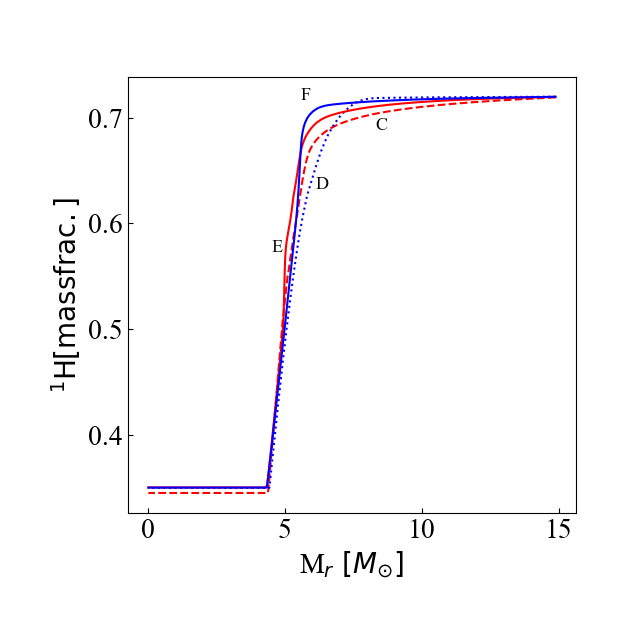}
    \includegraphics[width=0.33\textwidth]{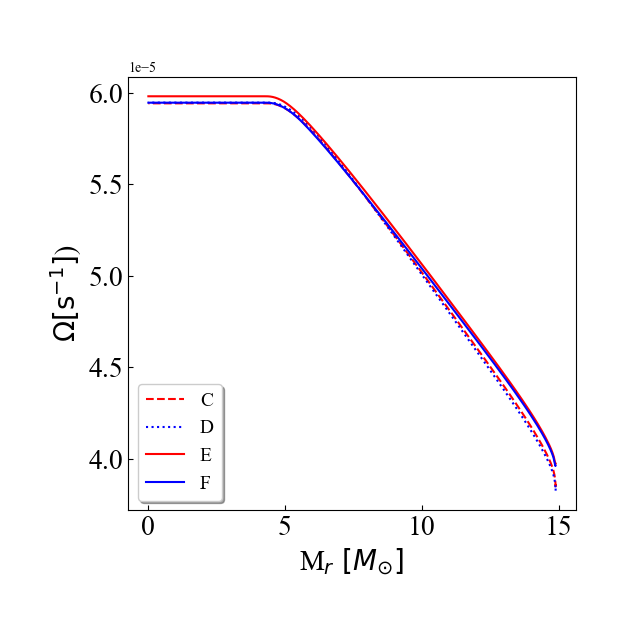}    
\caption{{\it Left panel:} Evolutionary tracks during the MS phase in the Hertzsprung-Russell diagram are shown
for models C, D, E and F (see Table~1) as a function of the Lagrangian mass for the 15 M$_\odot$ at $Z$=0.014 and with 
$V_{\rm ini}/V_{\rm crit}=0.4$. 
{\it Middle panel:} Abundance of Hydrogen in mass fraction ranging from the center to the outer envelope of the models when the central mass fraction of hydrogen $X_{\rm c}$=0.35.
{\it Right panel: } Variations of the angular velocity versus the mass coordinate for the same models as those shown in the middle panel.}
         \label{fig_Z14_CDEF_1}
   \end{figure*}

Figure~\ref{fig_Z14_CDEF_1} is the analog of Fig.~\ref{fig_Z14_MS} for models C, D, E and F. We see that models C and E  (using the $D_{\rm shear}$ of \citet{Maeder1997}, but different $D_{\rm h}$ expression (C the expression by
\citet{Maeder2003} and E, the expression of \citet{Mathis2004}) have very similar evolutionary tracks (see left panel). The model C shows a slightly smoother gradient of hydrogen than model E. 

Models F and D use the $D_{\rm shear}$ of \citet{TZ1997} with the two different $D_{\rm h}$. They present steeper chemical gradients above the convective core than models C and E that uses the $D_{\rm shear}$ by
\citet{Maeder1997}. This illustrates the fact that the choice $D_{\rm shear}$ impacts the degree of the chemical mixing. 

The angular velocity profiles between all these four models are nearly indistinguishable, illustrating the fact that the choice of $D_{\rm shear}$ and $D_{\rm h}$ has little impact on the angular momentum transport in the star. For the initial rotation considered here, the changes in the chemical mixing does not change so much the evolution as to indirectly impact the profile of $\Omega$.

\begin{figure*}
   \centering
    \includegraphics[width=0.33\textwidth]{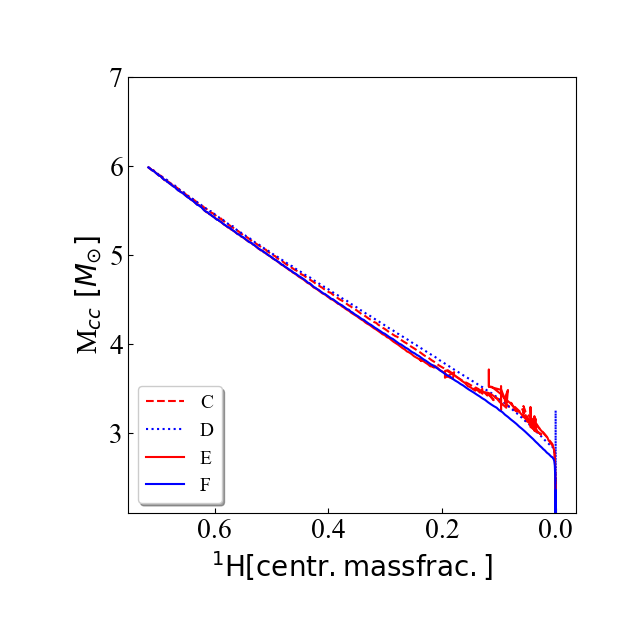}
    \includegraphics[width=0.33\textwidth]{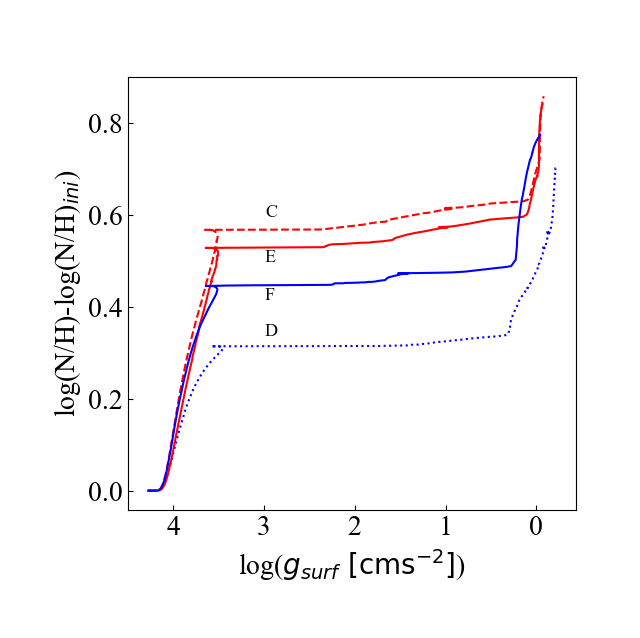}
    \includegraphics[width=0.33\textwidth]{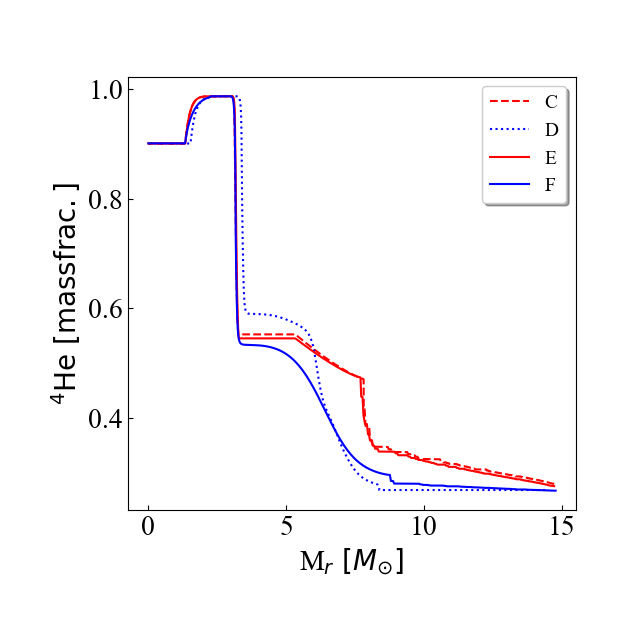}       
\caption{{\it Left panel:} Evolution of the convective core mass during the MS phase as a function of central mass of hydrogen.{\it Center panel:} The abundance ratio of nitrogen and hydrogen at the surface normalised to their initial values versus the surface gravity.{\it Right panel:} Profile of the helium mass fraction against the mass coordinate when the central helium fraction (Y$_c$ is 0.90). The letters `C',`D','E' and 'F' correspond to the models with different prescriptions as described in Table 1. }
         \label{fig_Z14_CDEF_222}
   \end{figure*}

The evolution of the convective core masses are compared in the right panel of Fig.~\ref{fig_Z14_CDEF_222}. They are not showing any striking differences.
On the other hand, we see that the surface enrichments are significantly different between these models (see the middle panel of Fig.~\ref{fig_Z14_CDEF_222}). Globally, the models using the D$_{\rm shear}$ of \citet{Maeder1997} (C and E) are more mixed than those using the D$_{\rm shear}$ of \citet{TZ1997} (D and F).

\begin{figure*}
   \centering
     \includegraphics[width=0.33\textwidth]{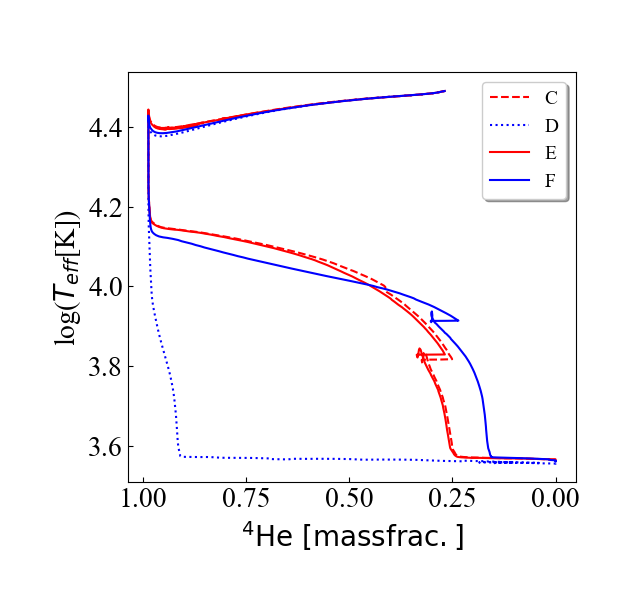}  \hfill\hfill
 
      \caption{Evolution of the effective temeprature as a function of the mass fraction of helium at the centre during the core He-burning phase for the models C, D, E and F.}
         \label{fig_cross}
   \end{figure*}

The abundances of helium as a function of the lagrangian mass coordinate when the mass fraction of helium at the centre is 0.9 is shown in the right panel of Fig.~\ref{fig_Z14_CDEF_222}. We can see that the profiles of models E and F are very similar to the one in model  C. Thus we expect that those models will spend a fraction of their core He-burning phase in a blue region of the HR diagram, which is indeed the case (see Fig.~\ref{fig_cross}). This supports the view that any helium enrichment in the H-burning shell tends to produce an evolution towards the red supergiant stage along a Kelvin-Helmholtz timescale.

\end{document}